\documentclass[letterpaper,twocolumn,10pt]{article}
\usepackage{usenix2019_v3}

\usepackage{tikz}
\usepackage{amsmath}

\usepackage{filecontents}

\usepackage{amsmath}
\usepackage{amssymb}
\usepackage{xstring}
\usepackage{comment}

\usepackage{xspace}
\usepackage{wrapfig}
\usepackage{booktabs}
\usepackage{float}
\usepackage{placeins}
\usepackage{url}
\usepackage{soul}

\usepackage{breakurl}
\usepackage{outlines}
\usepackage{pdfpages}
\usetikzlibrary{arrows.meta, positioning, backgrounds, calc, fit, shapes.geometric, decorations.pathmorphing, patterns
}
\usepackage{subcaption}
\usepackage{amssymb}
\usepackage{array,booktabs,multirow,tabularx}
\usepackage[dvipsnames]{xcolor}
\usepackage[textwidth=2cm,textsize=scriptsize]{todonotes}
\usepackage{multirow}
\usepackage{multicol}
\usepackage{array}

\usepackage{algorithm}
\usepackage{algpseudocode}
\algtext*{EndWhile}
\algtext*{EndFor}
\algtext*{EndIf}

\usepackage{filecontents}
\usepackage{makecell}

\usepackage{pgfplots, pgfplotstable}
\usepgfplotslibrary{groupplots}
\pgfplotsset{
    compat=1.18,
    unbounded coords=discard
}

\newcommand{\name}{\textsc{Agar}\xspace}

\newif\ifshowcomments
\showcommentstrue

\newcommand{\rps}{\ensuremath{\mathrm{rps}}}

\newcommand{\getreplicacloudlab}[3]{%
  \pgfplotstablegetrowsof{#1}%
  \pgfmathtruncatemacro{\lastrow}{\pgfplotsretval-1}%
  \def#3{--}%
  \foreach \r in {0,...,\lastrow}{%
    \pgfplotstablegetelem{\r}{autoscaler}\of{#1}%
    \edef\currentautoscaler{\pgfplotsretval}%
    \IfStrEq{\currentautoscaler}{#2}{%
      \pgfplotstablegetelem{\r}{replicas}\of{#1}%
      \xdef#3{\pgfplotsretval}%
      \breakforeach
    }{}%
  }%
}

\newcommand{\getreplica}[5]{%
  \pgfplotstablegetrowsof{#1}%
  \pgfmathtruncatemacro{\lastrow}{\pgfplotsretval-1}%
  \def#5{--}%
  \foreach \r in {0,...,\lastrow}{%
    \pgfplotstablegetelem{\r}{env}\of{#1}%
    \edef\currentenv{\pgfplotsretval}%
    \pgfplotstablegetelem{\r}{app}\of{#1}%
    \edef\currentapp{\pgfplotsretval}%
    \pgfplotstablegetelem{\r}{autoscaler}\of{#1}%
    \edef\currentautoscaler{\pgfplotsretval}%
    \IfStrEq{\currentenv}{#2}{%
      \IfStrEq{\currentapp}{#3}{%
        \IfStrEq{\currentautoscaler}{#4}{%
          \pgfplotstablegetelem{\r}{replicas}\of{#1}%
          \xdef#5{\pgfplotsretval}%
          \breakforeach
        }{}%
      }{}%
    }{}%
  }%
}

\begin{document}

\date{}

\title{\Large \bf \name: What-If Analysis for Microservice Applications}

\author{
{\rm Saarth Deshpande,
Neeraja J. Yadwadkar\thanks{Equal advisory contribution.},
Venkat Arun\hyperlink{Hfootnote.1}{\footnotemark[1]}}\\
The University of Texas at Austin
}

\maketitle

\begin{abstract}

Microservice operators must reason about how applications would behave under changes that have not yet occurred: different workloads, resource allocations, or SLO targets. Evaluating every alternative experimentally is expensive, but predicting such counterfactuals is difficult because microservices contain many interacting components. End-to-end models must learn behavior over a combinatorial space of conditions, but composing per-component predictors causes prediction errors to accumulate.

We address this challenge by changing what each component model must predict. Rather than predicting exact values, we learn simple, interpretable bounds that contain every observed behavior while remaining tight enough to support useful end-to-end conclusions. 
We build \name, a counterfactual performance analysis tool that composes these models to answer end-to-end SLO-feasibility questions. This allows us to formulate autoscaling as a counterfactual query: what is the smallest resource allocation under which an SLO is met?
Our evaluation demonstrates that across three microservice applications, \name's bounds are actionable: its derived autoscaler reduces CPU allocation by up to 36.8\% over state-of-the-art horizontal autoscalers while satisfying application SLOs. 
Further, \name is able to predict operational cost given workload traces, provide guidance on which service's startup time to optimize and localize performance bugs within each application. %

\end{abstract}

\section{Introduction}

Microservice operators constantly make decisions that can alter application performance: how many resources to allocate, how services are configured, what hardware to run them on~\cite{crisp}, and which service's performance to optimize~\cite{slowpoke}. A wrong choice can waste resources~\cite{autothrottle, harvestcontainers}, violate Service Level Objectives (SLOs)~\cite{firm-osdi, percs, cilantro}, or fail to address the actual bottleneck~\cite{sage, zeno}. 
However, the effect of a choice is often hard to know without deploying it. This is especially expensive in microservices, where local changes rarely stay local: changing one component can shift bottlenecks across the application. Performance can also change with the underlying environment, with different hardware surfacing failures that did not appear elsewhere. 
Because the impact of a choice must ultimately be established by observing application behavior over time~\cite{google-sre, google-sre-workbook-impl-slos}, evaluating each choice requires deploying and exercising the application under representative conditions, making exploration across many configurations expensive~\cite{oppertune}.

Operators need to know whether a proposed change will actually solve an observed performance problem, and what it will cost, without having to deploy the change.
This requires reasoning about how the application would behave under alternatives that have not been tried out yet. 
Would adding more resources restore SLO compliance or is the bottleneck elsewhere? 
Is the SLO unattainable under the current workload regardless of resource allocation?
How much cost could be saved if we relaxed the SLO by a certain amount? Would optimizing one component reduce the overall resource needs of the application?
These are counterfactual questions: they ask what would have happened under a change that was not actually made.

A growing body of systems work has explored counterfactual prediction for networks~\cite{veritas, causalsim}, storage~\cite{thesios, tiersizer}, and application behavior~\cite{sage, slowpoke, percs}. %
These systems show that learned models can predict behavior under decisions that were not actually exercised, but applying this idea to microservices exposes a fundamental scaling challenge. 
An end-to-end model must directly learn application behavior as a function of the workload, resource allocation, and an SLO target to effectively capture the joint behavior of many interacting microservices over a combinatorial space of conditions, many of which may never have been exercised.  %

A natural alternative is to decompose the application into per-component models. 
This narrows the behavior that each model must capture, but creates a different challenge: prediction errors compound as models are composed. 
With 50 component models, for example, naively guaranteeing 95\% confidence that the composition is correct would require each model to be correct roughly 99.9\% of the time.

The key is to \emph{change what we require from each component model}. Rather than requiring a model to predict a single value, such as CPU utilization, perfectly, we make its uncertainty explicit. For example, the model may predict that utilization lies between, say, 25--35\%. This allows models to achieve 100\% accuracy. Unfortunately, they may do so vacuously by always predicting a range of $0$ to $\infty$. In practice, %
however, our models achieve 100\% accuracy while remaining precise enough to be useful. This not only addresses the model-composability problem,
but also makes model errors and data drift easy to detect: the first observation that falls outside a predicted range signals that the model must be retrained.

Making uncertainty explicit also allows us to \emph{use simple models motivated by system intuition}. For example, we might model average CPU utilization as inversely proportional to the number of CPUs allocated. Such a model does not explicitly capture effects such as cross-VM interference or sudden bursts of garbage collection; instead, their observed effects are absorbed into the model's uncertainty bounds. More expressive models, including large neural models or LLM-based predictors, could attempt to capture these effects directly, but doing so introduces additional data requirements and complexity that are unnecessary for our goal. We do not need to predict behavior exactly; we need bounds that are tight enough to support useful counterfactual conclusions. Simple models can provide such bounds with less data, while remaining easier to interpret and compose across services.

We jointly fit the component models to minimize uncertainty in the end-to-end quantity predicted by their composition—for example, the average resource cost of meeting a given SLO—while requiring each model to fit all observed data. We prove that this optimization problem is NP-hard (Appendix~\ref{appendix:slo-algorithm}). We develop an efficient greedy algorithm with a $\log(n) + 1$ approximation guarantee for $n$ data points, and show that it performs well in practice.

Guided by these insights, we introduce \name, a counterfactual performance analysis tool to answer end-to-end SLO-feasibility questions. \name learns per-component models that fit \emph{every} observation, then composes them to reason about hypothetical changes in workload, resource allocation, and SLO targets. Using these models to identify the smallest resource allocation that can satisfy a target SLO directly yields an autoscaler. 
Across three microservice applications, we demonstrate that the bounds found by \name are actionable. Using these bounds, \name's autoscaler reduces CPU allocation by up to 36.8\% over state-of-the-art horizontal autoscalers while satisfying application SLOs. When the bounds fail to hold, they expose previously unknown performance problems in all three applications (see~\S\ref{subsec:debugger}).

\section{Motivation}

\pgfplotstableread[col sep=comma]{data/backpressure_hpa.csv}\hpadata
\pgfplotstableread[col sep=comma]{data/backpressure_whatif.csv}\whatifdata

\begin{figure}[!t]
\centering
\pgfplotsset{
  bp base/.style={
    width=\columnwidth, height=3.8cm,
    xlabel={Time (s)},
    xmin=12, xmax=1200,
    xtick={0,240,480,720,960},
    grid=both,
    grid style={line width=0.3pt, draw=gray!30},
    tick label style={font=\small},
    label style={font=\normalsize},
    legend cell align=left,
    thick,
  }
}

\begin{tikzpicture}
\begin{groupplot}[
  group style={
    group size=1 by 3,
    vertical sep=2.0cm,
  },
  bp base,
]

\nextgroupplot[
  ylabel={CPU utilization (\%)},
  label style={font=\small},
  ymin=0, ymax=205, ytick={50,100,150,200},
  legend style={font=\footnotesize, at={(0.55,0.97)}, anchor=north,
                draw=none, fill=white, fill opacity=0.75, text opacity=1, legend columns=1},
  after end axis/.code={
    \node[below=0.9cm, font=\small\bfseries, text width=\columnwidth, align=center]
      at (current axis.south)
      {\textbf{(a)} Backpressure keeps \texttt{frontend} CPU above the HPA threshold.};
  },
]
\addplot[red!5, fill=red!5, draw=none, forget plot]
  coordinates {(0,0)(0,50)(1200,50)(1200,0)} \closedcycle;
\addplot[green!80!black, thick]
  table[x=time, y expr=\thisrow{cpu_slow-service}, col sep=comma]
  {data/backpressure_hpa.csv};
\addlegendentry{\texttt{slow-service}}
\addplot[orange!80!black, thick]
  table[x=time, y expr=\thisrow{cpu_fast-service}, col sep=comma]
  {data/backpressure_hpa.csv};
\addlegendentry{\texttt{frontend}}
\addplot[black, dashed, thick]
  coordinates {(0,50)(1200,50)}
  node[pos=0.55, above, font=\footnotesize, text=black!70!black] {50\% threshold};
\node[below=0.9cm, font=\small\bfseries, text width=\linewidth, align=center]
  at (current axis.south)
  {};

\nextgroupplot[
  ylabel={Replicas},
  ymin=0, ymax=30, ytick={0,5,10,15,20,25,30},
  label style={font=\small},
  legend style={
    font=\footnotesize, 
    at={(0.57,0.53)}, anchor=north,
    draw=none, fill=white, fill opacity=1.0, text opacity=1,
    legend columns=2,
    column sep=0.5em,
  },
  after end axis/.code={
    \node[below=0.9cm, font=\small\bfseries, text width=\columnwidth, align=center]
      at (current axis.south)
      {\textbf{(b)} HPA scales \texttt{frontend} while \name scales the bottlenecked \texttt{slow-service}.};
  },
]
\addplot[red!80!black, dashed, thick, forget plot]
  table[x=time, y expr=\thisrow{replicas_fast-service}, col sep=comma]
  {data/backpressure_hpa.csv};
\addplot[red!80!black, thick, forget plot]
  table[x=time, y expr=\thisrow{replicas_slow-service}, col sep=comma]
  {data/backpressure_hpa.csv};
\addplot[blue!70!black, dashed, thick, forget plot]
  table[x=time, y expr=\thisrow{replicas_fast-service}, col sep=comma]
  {data/backpressure_whatif.csv};
\addplot[blue!70!black, thick, forget plot]
  table[x=time, y expr=\thisrow{replicas_slow-service}, col sep=comma]
  {data/backpressure_whatif.csv};

\addlegendimage{red!80!black, thick}
\addlegendentry{HPA}
\addlegendimage{black, dashed, thick}
\addlegendentry{\texttt{frontend}}
\addlegendimage{blue!70!black, thick}
\addlegendentry{\name}
\addlegendimage{black, thick}
\addlegendentry{\texttt{slow-service}}

\node[below=0.9cm, font=\small\bfseries, text width=\linewidth, align=center]
  at (current axis.south)
  {};

\nextgroupplot[
  height=4.2cm,
  label style={font=\small},
  ylabel={p95 latency (s)},
  ymin=0, ymax=7, ytick={0,2,4,6},
  legend style={
    font=\footnotesize,
    at={(0.88,0.99)}, anchor=north east,
    draw=none, fill=white, fill opacity=0.85, text opacity=1,
    legend columns=1,
  },
  after end axis/.code={
    \node[below=0.9cm, font=\small\bfseries, text width=\columnwidth, align=center]
      at (current axis.south)
      {\textbf{(c)} \name keeps p95 latency \mbox{below} 500\,ms by scaling \texttt{slow-service}.};
  },
]
\addplot[red!80!black, thick]
  table[x=time, y expr=\thisrow{process_p95}/1000, col sep=comma] {\hpadata};
\addlegendentry{HPA}
\addplot[blue!70!black, thick]
  table[x=time, y expr=\thisrow{process_p95}/1000, col sep=comma] {\whatifdata};
\addlegendentry{\name}

\end{groupplot}
\end{tikzpicture}

\caption{Backpressure case study. \textbf{(a)}~Under constant load, \texttt{frontend} workers busy-spin on the CPU while awaiting responses from \texttt{slow-service}, causing CPU utilization to remain above its HPA threshold longer. \textbf{(b)}~HPA consequently scales \texttt{frontend} to $\sim$18 replicas while adding only one \texttt{slow-service} replica, leaving the backend bottleneck unresolved. \textbf{(c)}~p95 latency climbs to $\sim$6\,s. \name instead identifies \texttt{slow-service} as the bottleneck and scales it out, keeping p95 below 500\,ms after its models converge.}
\label{fig:backpressure}
\end{figure}
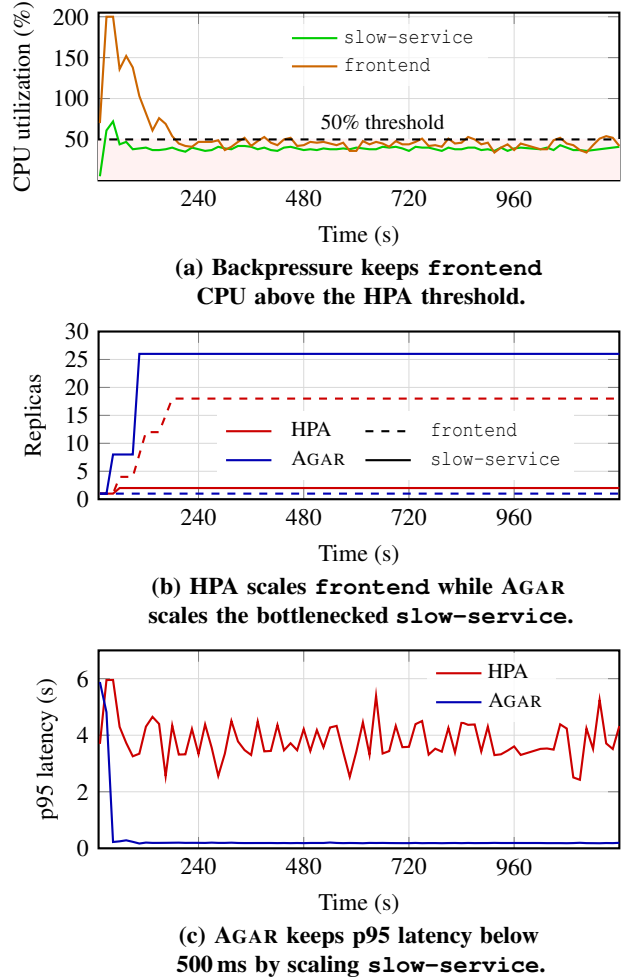

Running a microservice application requires making many performance-critical choices: how many resources to allocate to each service, how to configure scaling policies, what Service Level Objective (SLO) to target, and whether changes to the application or deployment are worth making. Answering these questions today often requires trying the corresponding configuration or change and measuring what happens~\cite{oppertune}. For applications with many interacting services, repeatedly doing so can require substantial experimentation.

Consider the amount of tuning required to even configure an application for one workload and SLO.
The Kubernetes Horizontal Pod Autoscaler (HPA)~\cite{k8s-hpa}, the most widely used autoscaler today~\cite{datadog-containers-serverless-2025}, requires operators to specify per-service scaling policies, such as what CPU or memory utilization should trigger scaling and how long that utilization must persist. For example, ``scale up if CPU utilization exceeds 60\% for more than 30\,s or memory utilization exceeds 80\% for 10\,s’’ for \emph{each service}.
Microservice applications may contain tens, hundreds, or even thousands of services~\cite{uber-failover-architecture}. Moreover, a configuration tuned for one set of conditions may perform poorly as those conditions change. This is not a one-time cost: Autopilot~\cite{autopilot} reports Google services whose changing resource demands required operators to manually retune CPU and memory allocations 8--12 times per month.

This tuning is further complicated by interactions between services. For example, when a backend service becomes a bottleneck, its frontend service must wait for responses, creating backpressure---delays that propagate from downstream to upstream services. The resulting backpressure can affect the frontend’s resource utilization even though scaling the backend is what relieves the bottleneck (Figure~\ref{fig:backpressure}). Such interactions make it difficult to determine which service should be scaled from utilization alone~\cite{deathstarbench}. %

Learning-based resource managers~\cite{erlang, firm-osdi, autothrottle, sinan-delimitrou, erms, percs} can automate part of this tuning, using application measurements to guide per-service resource allocation through horizontal or vertical scaling and reduce resource use while targeting end-to-end latency SLOs. But operators often need answers to questions that go beyond finding a configuration for the application as it runs today. Suppose an engineer could reduce the startup time of one service. Would that change meaningfully reduce resource cost or improve SLO satisfaction? Suppose the traffic mix shifts toward one endpoint\footnote{An endpoint is a distinct externally visible API or RPC call of an application, each with its own latency SLO.}. Will the current configuration still suffice, and if not, how should resource allocation change? Suppose the SLO is tightened or relaxed. What would that do to the resources required to meet the new SLO? Evaluating every such possibility by deploying it and re-tuning the application turns each question into another experiment. %

Consider what happens when the workload changes. A shift in traffic mix changes how requests flow through the application's services and can move bottlenecks from one component to another. Before we can answer any counterfactual question, the system must be re-tuned to the new workload. HPA~\cite{k8s-hpa} leaves this entirely up to the developer. Erlang~\cite{erlang} helps, but requires a new instance of the application that is \emph{not} serving user traffic to experiment on and train whenever the workload moves outside its training parameters. Erlang therefore still relies on active experimentation to learn how the application behaves under new conditions.

In contrast, \name can learn passively from observed data and reuse what it has learned across conditions. The key is that \name models simple, per-service relationships and composes them to answer end-to-end questions. For example, a change in traffic mix may alter how much load reaches a service without changing how its utilization responds to load and resource allocation. \name can therefore reuse the models that map each service's load to its utilization to determine SLO satisfaction under the new workload rather than retraining the application from scratch. When such reuse is no longer valid, \name's ``always accurate'' models make this immediately detectable: the first observation that violates a learned relation identifies a model that must be retrained. Such violations can occur when the deployment environment changes. For example, when we moved Online Boutique~\cite{online-boutique-github} and TeaStore~\cite{teastore} from AWS to CloudLab~\cite{cloudlab}, CloudLab's slower storage exposed performance problems that had not appeared on AWS, invalidating relationships learned from the earlier deployment (see~\S\ref{subsec:debugger}). In our experiments, such retraining was needed only when an environment change exposed previously unseen performance bugs. This allows \name to make predictions far from prior experiments without requiring a new tuning run for every workload, SLO target, or candidate change.

Other systems reduce the amount of experimentation in a different way: they restrict the dimensionality or locality of the problem being learned. Galileo~\cite{percs}, a vertical autoscaler\footnote{Vertical autoscaling manages resources allocated to a single VM. Horizontal autoscaling, our focus, allocates the number of VMs.}, deliberately models only the effect of \emph{local changes} to resource allocations. This allows it to make linearizing assumptions and use compressed sensing to reduce the number of experiments needed to identify the bottleneck. AutoThrottle~\cite{autothrottle} similarly reduces the search space by dividing services into two groups and tuning one throttling target per group to guide per-service CPU allocation, while Slowpoke~\cite{slowpoke} answers counterfactual questions about optimizing a \emph{given} service by \emph{actively} slowing down every other service. These systems still rely on active experiments, but keep the learning problem tractable by narrowing the space over which they must reason.

These observations suggest a different point in the design space: a system should reuse simple component-level relationships across workloads and configurations, compose them to answer end-to-end counterfactual questions, and detect when those relationships no longer hold rather than requiring active experimentation whenever conditions change. \name is designed around these principles. We next describe how \name represents component behavior, learns these models from passive observations, and composes them to answer counterfactual queries.

\section{\name}

\name is a counterfactual performance analysis tool for microservice applications that answers SLO-feasibility questions under hypothetical changes to workload, resource allocation, and SLO targets. To answer these questions, \name fits models that relate workload and resource allocation to each microservice’s resource utilization, and that utilization to end-to-end SLO satisfaction.

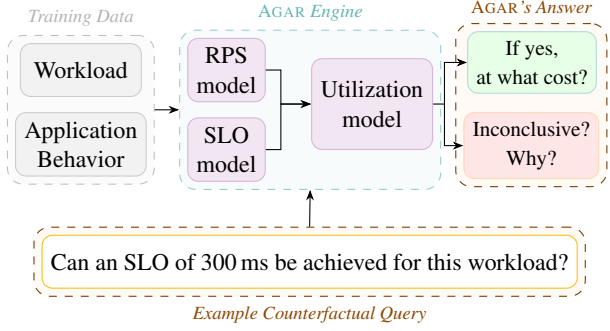
\begin{figure}[!t]
\centering
\begin{tikzpicture}[
    node distance=0.4cm and 0.5cm,
    box/.style={rectangle, rounded corners=4pt, draw, minimum width=1.6cm, minimum height=0.7cm, align=center, font=\small},
    graybox/.style={box, fill=gray!10, draw=gray!40},
    purplebox/.style={box, fill=violet!10, draw=violet!30},
    tealbox/.style={box, fill=green!10, draw=teal!30},
    coralbox/.style={box, fill=red!10, draw=orange!30},
    querybox/.style={box, fill=white!10, draw=yellow!45!orange},
    arr/.style={-{Stealth[length=4pt]}, thin},
    lbl/.style={font=\scriptsize},
]

\node[purplebox, minimum width=1.0cm] (rps) {RPS\\model};
\node[purplebox, minimum width=1.0cm, below=0.2cm of rps]
    (slomodel) {SLO\\model};

\node[graybox, left=0.6cm of rps]
    (workload) {Workload};
\node[graybox, below=0.2cm of workload]
    (behavior) {Application\\Behavior};

\node[purplebox, right=0.6cm of rps, yshift=-0.45cm,
      minimum height=1.2cm] (ms) {Utilization\\model};

\node[tealbox, right=0.4cm of ms, yshift=0.55cm, align=center, xshift=0.05cm] (cost)
    {\footnotesize If yes,\\\footnotesize at what cost?};
\node[coralbox, right=0.4cm of ms, yshift=-0.55cm, align=center] (bottleneck)
    {\footnotesize Inconclusive?\\ \footnotesize Why?};

\node[querybox, below=0.7cm of slomodel, xshift=1.1cm] (query)
    {Can an SLO of 300\,ms be achieved for this workload?};

\begin{pgfonlayer}{background}
    \node[
        draw=gray!50, dashed, rounded corners=6pt, fill=gray!3,
        inner sep=0.1cm,
        fit=(workload)(behavior),
        label={[font=\scriptsize\itshape, text=gray!60]above:Training Data},
    ] (observations) {};

    \node[
        draw=teal!40, dashed, rounded corners=6pt, fill=teal!3,
        inner sep=0.1cm,
        fit=(rps)(slomodel)(ms),
        label={[font=\scriptsize\itshape, text=teal!60]above:\name{} Engine},
    ] (engine) {};

    \node[
        draw=orange!45!black, dashed, rounded corners=6pt, fill=orange!4,
        inner sep=0.1cm,
        fit=(cost)(bottleneck),
        label={[font=\scriptsize\itshape, text=orange!55!black]above:\name's Answer},
    ] (answer) {};

    \node[
        draw=orange!45!black, dashed, rounded corners=6pt, fill=orange!4,
        inner sep=0.1cm,
        fit=(query),
    ] (question) {};
\end{pgfonlayer}

\node[
    font=\scriptsize\itshape,
    text=orange!55!black,
    align=center,
    below=0cm of question
] {Example Counterfactual Query};

\draw[arr] (observations.east) -- (engine.west);

\draw[arr] (question.north) -- (engine.south);

\draw[arr] (rps.east) -- ++(0.2,0) |- (ms.west);

\draw[arr] (slomodel.east) -- ++(0.2,0) |- (ms.west);

\coordinate (split) at ($(ms.east)+(0.15,0)$);
\draw (ms.east) -- (split);
\draw[arr] (split) |- (cost.west);
\draw[arr] (split) |- (bottleneck.west);

\end{tikzpicture}
\caption{\name system overview.}
\label{fig:system-overview}
\end{figure}

\subsection{\name Architecture}
Figure~\ref{fig:system-overview} shows the architecture of \name.  Users interact with \name through a fixed set of counterfactual queries: how many resources are needed to meet a given SLO; how that resource requirement changes for a different workload or SLO; and how changing a service property, such as its startup time, affects the resources needed to meet the SLO. To answer these questions, \name uses per-endpoint request rates and latencies and per-service CPU and memory utilization collected from previous executions. \name uses these measurements to fit the models described in the following sections. Together, these models comprise the \name Engine.

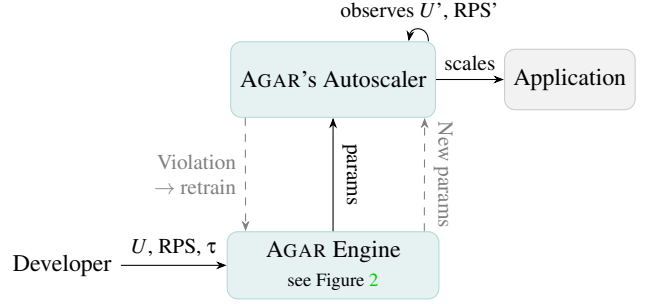
\begin{figure}[!t]
\centering
\begin{tikzpicture}[
    node distance=0.6cm and 1.0cm,
    box/.style={rectangle, rounded corners=4pt, draw, minimum width=2.5cm, minimum height=0.8cm, align=center, font=\small},
    graybox/.style={box, fill=gray!10, draw=gray!40},
    tealbox/.style={box, fill=teal!10, draw=teal!30},
    arr/.style={-{Stealth[length=4pt]}, thin},
    darr/.style={-{Stealth[length=4pt]}, thin, dashed, gray},
    lbl/.style={font=\footnotesize},
]

\node[tealbox, minimum width=2.5cm, minimum height=1.0cm] (autoscaler)
    {\name's Autoscaler};

\node[graybox, minimum width=1.7cm, right=0.9cm of autoscaler] (app) {Application};

\node[tealbox, below=1.5cm of autoscaler, minimum width=2.8cm] (engine)
    {\name Engine \\ \scriptsize see Figure~\ref{fig:system-overview}};

\node[left=1.4cm of engine, font=\small] (dev) {Developer};

\draw[arr] (dev.east) -- (engine.west)
    node[midway, above, lbl] {$U$, RPS, $\tau$};

\draw[arr] (engine.north) -- (autoscaler.south)
    node[midway, right, lbl, rotate=-90, anchor=south] {params};

\draw[arr] ($(autoscaler.north east)+(-0.1,0)$)
    to[out=90, in=80, looseness=2] 
    node[above, lbl] {observes $U$', RPS'}
    ($(autoscaler.north)+(1.0,0)$);

\draw[arr] (autoscaler.east) -- (app.west)
    node[midway, above, lbl] {scales};

\draw[darr] ($(autoscaler.south west)+(0.2,0)$) -- ($(engine.north west)+(0.25,0)$)
    node[midway, left, lbl, align=right] {Violation \\ $\rightarrow$ retrain};

\draw[darr] ($(engine.north east)+(-0.2,0)$) -- ($(autoscaler.south east)+(-0.15,0)$)
    node[midway, right, lbl, rotate=-90, anchor=south] {New params};

\end{tikzpicture}
\caption{\name autoscaler control loop. The developer provides the initial observed utilization $U$, RPS, latencies and SLO target $\tau$ to fit the \name Engine's models. At runtime, the autoscaler periodically observes $U$' and RPS' from the cluster, and uses the learned parameters $(\alpha(\cdot), \beta, \epsilon, \tau, \Delta T(\cdot))$ to answer the counterfactual query: \emph{given $U'$ and $RPS'$, what is the minimum replica count that satisfies $\tau$?} When a model bound is exceeded or an SLO violation occurs, it retrains by calling the \name Engine to update its parameters.}
\label{fig:autoscaler-loop}
\end{figure}

Autoscaling is one such query issued repeatedly: given predicted request rates and a target SLO, \name computes how many replicas each service needs to meet the SLO. As the workload changes, \name repeats this query to determine the next resource allocation. During autoscaling, runtime monitors check new request-rate, utilization, and latency observations against the learned relationships and trigger retraining when an observation falls outside the corresponding model bounds, as shown in Figure~\ref{fig:autoscaler-loop}.

\subsection{Models and Modeling Philosophy}\label{sec:modeling-philosophy}

We describe application behavior under different workloads and resource allocations using three models: (1) the Utilization Model that relates per-endpoint request rates and each service's replica count to its CPU and memory utilization; (2) the SLO Model that relates these utilizations 
to whether each endpoint's SLO is met; and (3) the RPS Model that relates past request rates to future demand. Figure~\ref{fig:system-overview} shows how these models compose within the \name Engine.

Recall that these are \emph{relations}, not exact predictions. For example, rather than predicting that a particular workload will result in exactly 60\% CPU utilization, we might say that its CPU utilization should fall between 50\% and 70\%. This distinction lets us learn relations that hold for 100\% of the training observations.

\paragraph{Utilization Model.} To retain this ability to fit all observed behavior, 
we deliberately make weak assumptions about application behavior. To model the utilization $U_i$ of service $i$, we assume that it comprises three components: a workload-dependent component, $\sum_j \alpha_{i,j}$, divided evenly across $P_i$ replicas, where $\alpha_{i,j}$ is the utilization contributed by endpoint $j$ to service $i$ and increases monotonically with its request rate (RPS), a workload-independent component, $\beta_i \ge 0$, and bounded variation $\epsilon_i \ge 0$. Specifically, we assume the following relation:

\begin{equation}
    U_i \in \frac{\sum_j \alpha_{i,j}(\rps_j)}{P_i} + \beta_i \pm \epsilon_i
    \label{eqn:utilization-constraint}
\end{equation}

Importantly, we make no assumption about the shape of $\alpha_{i,j}$ beyond monotonicity: different endpoints and services can therefore exhibit different relationships between request rate and utilization. We use this relation for both CPU and memory utilization. 

Different services exercise this flexibility in different ways. Some services can retain memory in heaps and caches even when request rate is low or zero, resulting in a substantial workload-independent component captured by $\beta_i$. Garbage collection, for example, can introduce large, irregular changes in CPU and memory utilization, which manifest as a larger $\epsilon_i$. Thus, the same relation can capture qualitatively different utilization behavior across services without changing its structure.

\paragraph{SLO Model.} We next apply the same philosophy to SLO satisfaction: instead of predicting latency, we identify utilization bounds within which an SLO is satisfied. Our core assumption is that there exists a utilization below which a target SLO is always satisfied. Above these bounds, observations may either satisfy or violate the SLO, and we make no claim about them. Figure~\ref{fig:slo-model-example} illustrates the intuition behind constructing these bounds. Any bounds that exclude all observed SLO violations optimistically treat the enclosed region as SLO-satisfying, even if no observations fall inside it. There can be many such bounds, and we describe how \name selects among them in~\S\ref{subsec:joint-optimization}.

\paragraph{RPS Model.} Finally, to account for service startup delays ($\Delta C_j$), \name predicts future request rates so that replicas can be provisioned before the demand arrives. Like HPA, which reacts to the maximum utilization over its \texttt{behavior.scaleUp.stabilizationWindowSeconds}~\cite{k8s-hpa-stabilization-window} window, we use the maximum request rate over the previous $t$ seconds. The simplicity is intentional: \name requires a prediction of future demand, but not a particular prediction method. More sophisticated predictors can therefore be substituted directly. Despite its simplicity, future request rate exceeds our prediction in at most $\frac{1}{E}\%$ observations, where $E$ is the number of application endpoints.

\definecolor{thresholdBlue}{HTML}{1676B8}
\definecolor{violationRed}{HTML}{D62728}

\pgfplotsset{
  threshold axis/.style={
    width=2.5cm, height=2.5cm,
    scale only axis,
    xmin=0, xmax=101, ymin=0, ymax=101,
    axis lines=left,
    axis line style={-},
    xtick={0,20,60,99},
    ytick={0,20,60,99},
    tick align=outside,
    tick label style={font=\small},
    xlabel={Service A utilization (\%)},
    ylabel={Service B utilization (\%)},
    label style={font=\small},
    clip=false,
  },
  observation/.style={
    only marks,
    mark=x,
    mark size=3.2pt,
    draw=violationRed,
    mark options={line width=1.2pt},
  },
}

\newcommand{\ThresholdBox}[2]{%
  \fill[thresholdBlue,fill opacity=0.10]
    (axis cs:0,0) rectangle (axis cs:#1,#2);
  \draw[thresholdBlue,dashed,line width=1pt]
    (axis cs:0,#2) -- (axis cs:#1,#2)
                    -- (axis cs:#1,0);
}

\newcommand{\PlotPoints}[1]{%
  \addplot[observation]
    coordinates {(60,20) (20,60)};
  \addplot[observation,opacity=#1]
    coordinates {(80,80)};
}

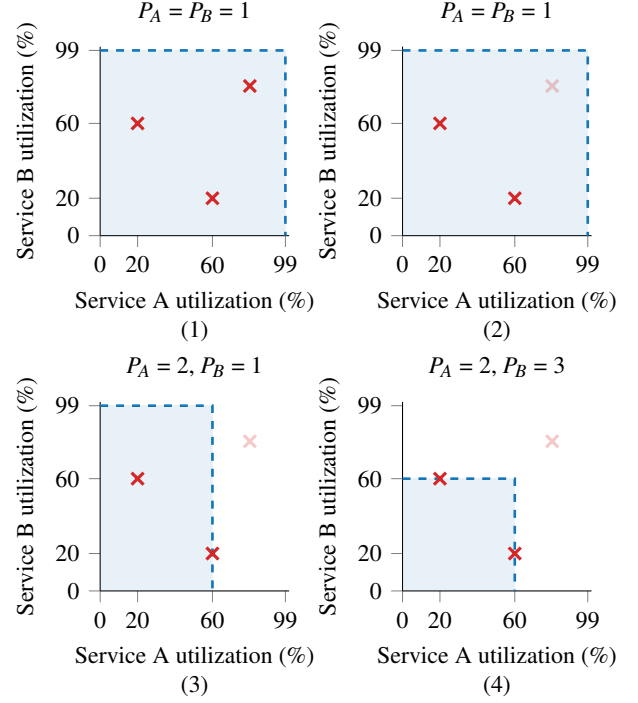
\begin{figure}[t]
\centering
\begin{tikzpicture}

\begin{groupplot}[
  threshold axis,
  group style={
    group size=2 by 2,
    horizontal sep=1.5cm,
    vertical sep=2.2cm,
  },
]

\nextgroupplot
\ThresholdBox{99}{99}
\PlotPoints{1}
\node[anchor=south,font=\small]
  at (axis description cs:0.5,1.08) {$P_A$ = $P_B$ = 1};
\node[anchor=north,font=\small]
  at (axis description cs:0.5,-0.40) {(1)};

\nextgroupplot
\ThresholdBox{99}{99}
\PlotPoints{0.25}
\node[anchor=south,font=\small]
  at (axis description cs:0.5,1.08) {$P_A$ = $P_B$ = 1};
\node[anchor=north,font=\small]
  at (axis description cs:0.5,-0.40) {(2)};

\nextgroupplot
\ThresholdBox{60}{99}
\PlotPoints{0.25}
\node[anchor=south,font=\small]
  at (axis description cs:0.5,1.08) {$P_A$ = 2, $P_B$ = 1};
\node[anchor=north,font=\small]
  at (axis description cs:0.5,-0.40) {(3)};

\nextgroupplot
\ThresholdBox{60}{60}
\PlotPoints{0.25}
\node[anchor=south,font=\small]
  at (axis description cs:0.5,1.08) {$P_A$ = 2, $P_B$ = 3};
\node[anchor=north,font=\small]
  at (axis description cs:0.5,-0.40) {(4)};

\end{groupplot}
\end{tikzpicture}

\caption{Illustration of selecting utilization bounds for SLO satisfaction. Crosses denote SLO violations observed at corresponding service utilizations, the shaded region denotes utilizations predicted to satisfy the SLO, and $P_A$ and $P_B$ denote the replica counts for services $A$ and $B$ under the corresponding utilization bounds. We (1) optimistically start with 99\% utilization bounds, (2) discard violations dominated by another violation, since excluding the latter also excludes them, and (3) tighten utilization bounds at the lowest additional replica cost per newly excluded SLO violation. (4) We repeat until all observed violations lie outside the shaded region.}\label{fig:slo-model-example}
\end{figure}

\subsection{Composing and jointly optimizing the \mbox{models}}\label{subsec:joint-optimization}

The models above work together to answer \name's counterfactual queries. Consider determining how many replicas each service needs to meet an SLO $\tau$. \name first predicts the request rate at the end of each service’s startup delay. Given these predicted request rates, the utilization model bounds each service’s CPU and memory utilization for a candidate replica count. The SLO model then determines whether these utilization bounds are low enough to satisfy $\tau$.

Let $T_i$ denote an SLO-safe utilization threshold for service $i$, and let $\rps^{\mathrm{max}}_j$ denote the predicted request rate for endpoint $j$. For the SLO to be satisfied, the upper bound on service $i$’s utilization must remain below $T_i$. Applying this condition to Equation~\ref{eqn:utilization-constraint} gives the required replica count:

\begin{equation}
\label{eqn:num-pods}
P_i = \frac{\sum_j \alpha_{i,j}(\rps^{\mathrm{max}}_j)}{T_{i} - \beta_{i} - \epsilon_{i}}
\end{equation}

Once the SLO model produces $T_i$, \name fits the utilization model to minimize the replica count in Equation~\ref{eqn:num-pods}. The resulting utilization-model optimization is fractional, so we use the Charnes–Cooper transformation~\cite{charnes-cooper-1962} to convert it into a sequence of linear programs to find the optimal solution in polynomial time (Appendix~\ref{appendix:lp-fitting}). 

The harder problem is choosing the SLO-safe utilization thresholds. Many choices can exclude all observed SLO violations, but they can require different numbers of replicas. For example, suppose an SLO violation occurs when one service has high CPU utilization and another has high memory utilization. Lowering either service’s corresponding threshold can exclude that violation, but each choice can imply a different replica cost through Equation~\ref{eqn:num-pods}. Thus, there can be multiple ways to exclude each violation, with different costs.

Choosing the optimal set of thresholds is NP-hard (proof in Appendix~\ref{appendix:slo-algorithm}). However, we develop a simple greedy algorithm that we prove finds a solution within a factor of $\log(n) + 1$ of the optimal number of replicas needed, where $n$ is the number of SLO violations in training data. 

\name's greedy algorithm is inspired by the standard greedy set-cover approximation~\cite{chvatal1979greedy}. \name initially sets each utilization bound to its largest valid candidate, up to 99\%. Each service’s initial replica budget is the maximum replica requirement across its utilization dimensions. At each step, \name updates one service’s replica budget, choosing the move with the smallest additional replica cost per newly excluded violation. It fits utilization models at candidate thresholds on demand to determine this move. Once all violations are covered, \name selects the final thresholds for each updated service. For each utilization dimension, it selects the threshold that excludes the most retained SLO violations within the service’s budget, choosing the largest valid threshold when several exclude the same violations.

Importantly, this optimization does not change the assumptions made by the individual models. It selects, among models that fit all observed behavior, those whose composition gives the most useful answer to the end-to-end query.

\vspace{-2mm}

\section{Implementation}\label{subsec:implementation}

\begin{figure}[t]
\centering
\begin{tikzpicture}[
    box/.style={rectangle, rounded corners=4pt, draw, minimum width=2.2cm,
                minimum height=0.65cm, align=center, font=\small},
    graybox/.style={box, fill=gray!10, draw=gray!40},
    tealbox/.style={box, fill=teal!10, draw=teal!30},
    dashedbox/.style={rectangle, rounded corners=6pt, draw=gray!50, dashed,
                      inner sep=6pt},
    arr/.style={-{Stealth[length=4pt]}, thin},
    lbl/.style={font=\scriptsize},
]

\node[tealbox, minimum width=2.2cm] (wi) {\name's Autoscaler};
\node[lbl, left=0.9cm of wi] (rps) {\small RPS};
\node[graybox, below=0.8cm of wi] (hpa) {HPA Controller};
\node[graybox, right=1.6cm of hpa] (metrics) {Metrics Server};
\node[graybox, below=0.7cm of hpa] (deploy) {Deployment};

\begin{scope}[on background layer]
  \node[dashedbox, fit=(hpa)(metrics)(deploy),
        label={[font=\scriptsize\color{gray}]north west:Kubernetes}] (k8s) {};
\end{scope}

\draw[arr] (rps.east) -- (wi.west);

\draw[arr] (wi.south) -- (hpa.north)
    node[midway, left, lbl] {$P_t$};

\draw[arr] (metrics.west) -- (hpa.east)
    node[midway, above, lbl] {Utilizations};

\draw[arr] (metrics.north) -- ++(0,0.9) |- (wi.east)
    node[pos=0.35, below, xshift=-1.0cm, lbl] {\texttt{kubectl top}};

\draw[arr] (hpa.south) -- (deploy.north);

\end{tikzpicture}
\caption{%
  \name's Autoscaler sits outside the Kubernetes control plane (dashed region). At each tick, it reads per-service utilization via \texttt{kubectl top} from the Metrics Server and patches the service's HPA object to $P_t$, causing HPA to adjust the service's Deployment to $P_t$ pods. %
}
\label{fig:k8s-stack}
\end{figure}
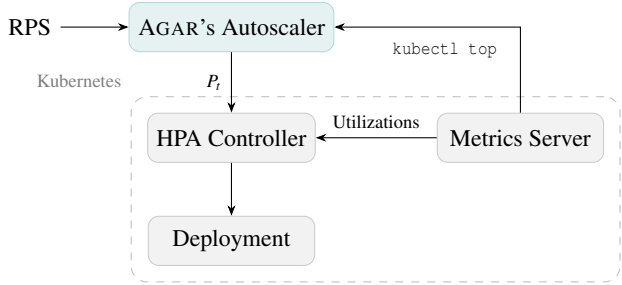

\name's Autoscaler runs as a per-service control loop on a 15-second tick, matching HPA's default scrape granularity. Like HPA, it operates per-service and reads cluster metrics via the Kubernetes metrics API. \name's Autoscaler piggybacks on the HPA (Figure~\ref{fig:k8s-stack}): at each tick, it evaluates Equation~\ref{eqn:num-pods} to obtain the minimum replica count $P_t$ for service $s$ and patches the existing HPA object by pinning both \texttt{minReplicas} and \texttt{maxReplicas} to exactly $P_t$, while leaving all underlying Kubernetes pod management intact. 
To compute $P_t$, \name's Autoscaler evaluates Equation~\ref{eqn:num-pods} using the current RPS observed from the cluster and the parameters learned offline by the \name Engine. If no feasible model exists for a service, \name's Autoscaler falls back to the  HPA. However, we have not encountered this case in practice.

Because the parameters are fit offline on historical data, \name's Autoscaler monitors the application's live observability data for conditions that indicate the model no longer reflects the application's current behavior and triggers a retrain when they arise, off the critical path. Across all experiments in~\S\ref{subsec:autoscaler-eval}, full retraining was required at most once while utilization bounds were updated throughout each run.

\textbf{Utilization-bound violation.} After each tick, \name's Autoscaler checks whether the observed utilization $U_{t}$ lies within the bounds predicted by Equation~\ref{eqn:utilization-constraint}. A violation indicates that the learned parameters no longer describe the service's behavior — for example, because the workload has shifted or pods have migrated to different nodes. \name's Autoscaler responds by retraining the utilization model in the background, updating its parameters, and reloading them into the live control loop. A per-metric cooldown of 60 seconds prevents retrain storms when violations are transient.

\textbf{SLO violation.} Every 60 seconds, a separate monitor checks whether any end-to-end latency measurement exceeds the SLO target $\tau$. An SLO violation for \emph{any} endpoint leads to a full-application retrain: \name's Autoscaler calls the \name Engine (Figure~\ref{fig:autoscaler-loop}) to re-fit all service models from the accumulated data and pushes updated thresholds to the deployed application. Full retrains are rate-limited to one per five minutes; if a retrain is already in progress, the request is dropped and the next monitoring cycle re-evaluates. In both cases, the retrain writes updated parameters to disk; per-service utilization monitors detect the change and reload before the next tick. 

\name's implementation separates offline model fitting from online autoscaling. We implement model fitting in Python using PuLP~\cite{pulp} with the CBC solver. For online operation, each service runs its own autoscale loop in a separate process. Utilization-bound retrains for a single service are submitted asynchronously to a single-worker executor, so retrains are serialized but never block the tick loop. Full retrains are guarded by a cross-process lock so that a slow full retrain neither races with a concurrent single-service retrain nor starves the autoscale loops waiting on updated parameters. A shared tick-coordinator thread snapshots RPS and latency readings atomically at each 15-second boundary, ensuring all per-service loops operate on a consistent view of the cluster state.

As \name's Autoscaler operates, it continuously appends observations to the training set. As the training set grows to cover more operating conditions, the model fits tighten and bound violations become increasingly rare, and the autoscaler settles into a steady state where $P_t$ changes only in response to genuine shifts in demand.

\vspace{-2mm}

\section{Evaluation}

Our evaluation demonstrates \name's use across autoscaling, performance diagnosis, and counterfactual changes to workloads, resource allocation, and SLO targets.

\subsection{Experimental Setup}

\paragraph{Cluster.}
We run experiments on two otherwise identical AWS EC2 clusters, each comprising 15 \texttt{t3.2xlarge} instances\footnote{Standard \texttt{t3} instances allow noisy-neighbor CPU steal to manifest, unlike fixed-performance instances whose vCPUs are pinned to physical cores~\cite{aws-nitro-side-channels}. These instances also experience throttling when their CPU credits are exhausted~\cite{aws-burstable-credits}.}: one control-plane instance and 14 worker instances running Kubernetes deployed with \texttt{kubeadm}. Application pods are scheduled only on the worker instances; \name's Autoscaler runs on the control-plane instance. One cluster runs on an EC2 Dedicated Host~\cite{aws-dedicated-hosts}, avoiding host-level interference from colocated instances, while the other uses shared tenancy and is exposed to such interference. The dedicated and shared clusters also let us evaluate \name under host-level interference. \name does not model noisy neighbors explicitly; instead, their effects appear in the utilization relationships learned from each environment. Short-term interference broadens the range of utilization observed for a given workload, while sustained interference shifts the learned baseline or workload-utilization relationship. We quantify these parameter changes in Appendix~\ref{model-param-interference-changes}. 

\vspace{-1mm}

\paragraph{Benchmarks.}
We evaluate on three microservice applications: TeaStore~\cite{teastore} (7 services), Online Boutique~\cite{online-boutique-github} (11 services), and Social Network~\cite{deathstarbench} (20 services). For Social Network, we exclude MongoDB instances from the autoscaled service set. Although these components are deployed as Kubernetes \texttt{Deployment} objects rather than \texttt{StatefulSet}\footnote{A \texttt{StatefulSet} assigns each pod a stable network identity and persistent storage, making it the appropriate abstraction for stateful workloads; a \texttt{Deployment} treats all pods as interchangeable.} objects, the deployment does not support sharding, so multiple replicas would contend for the same persistent volume. We also exclude the Jaeger distributed tracing service. For Online Boutique, we replace the bundled load generator with an external Locust~\cite{locust} instance.

Kubernetes configurations contain a guaranteed CPU and memory allocation (requests), and a maximum allocation that may be reached if resources are available (limits). For Online Boutique, each replica is configured with CPU requests and limits of 1000m and 2000m respectively, and memory requests and limits of 64 MiB and 128 MiB. The exception is \texttt{redis-cart}, which retains its default configuration: CPU requests and limits of 70m and 125m, and memory requests and limits of 200 MiB and 256 MiB. TeaStore uses the same CPU configuration, but sets both memory requests and limits to 2 GiB. Social Network uses CPU requests and limits of 500m and 1000m respectively, and memory requests and limits of 128 MiB and 256 MiB. 

\paragraph{Workloads.}
For each application, we derive replay workloads from Alibaba production traces~\cite{alibaba-2021}, and use the HTTP-entrypoint microservice call rate (MCR) as the workload signal. These HTTP entrypoints provide the most direct measure of offered load. Using call-rate measurements from internal service interactions would instead make the workload signal depend on application-specific implementation details rather than externally imposed demand. Since the original request-rate data is available at 60\,s granularity, we linearly interpolate it to 15\,s resolution for replay. We assign one such workload trace to each application endpoint and rescale the resulting demand to match the capacity of the target deployment. Locust is configured with 100 users and a ramp-up of 1 user/s for TeaStore and Online Boutique, and with 1000 users and a ramp-up of 100 users/s for Social Network.

\paragraph{Baselines.} We compare \name against Erlang~\cite{erlang}, a bandit-learning horizontal autoscaler, and against the Kubernetes Horizontal Pod Autoscaler (HPA)~\cite{k8s-hpa} configured with hand-tuned utilization thresholds and scaling policies. Most prior research focuses on \emph{vertical} autoscaling\cite{sinan-delimitrou,percs,autothrottle}. We therefore use Erlang as our research baseline for horizontal autoscaling. Our HPA configurations are the best we are aware of after extensive experimentation, obtained by iteratively adjusting thresholds and observing scaling behavior under load. Many production horizontal autoscalers, like HPA, leave users to tune scaling targets for their applications~\cite{knative-kpa,gcp-autoscaler, aws-autoscaling,gke-metric-autoscaling}.

\paragraph{Training and testing.} For each application, \name's models are fit on the first 6 of 12 hours of the Alibaba-derived workload~\cite{alibaba-2021}, and Erlang is trained on the RPS contexts of the same duration. Autoscaler performance is evaluated on the following 2-hour period of each trace; representative workload segments are shown in Appendix~\ref{appendix:workloads}. While the models converge within 4 minutes for Online Boutique and Social Network and 12 minutes for TeaStore, we report results for the entire duration, including the period before convergence. Across all experiments, full retraining was required at most once while utilization bounds were updated throughout each run.

\subsection{\name as an Autoscaler}\label{subsec:autoscaler-eval}

\definecolor{dedicatedColor}{HTML}{3B82F6}
\definecolor{sharedColor}{HTML}{F97316}

\pgfplotstableread[col sep=comma]
    {data/autoscaler-numbers.csv}\latencydata

\pgfdeclareplotmark{filled plus}{%
    \pgfpathrectangle
        {\pgfpoint{-\pgfplotmarksize}{-.333333\pgfplotmarksize}}
        {\pgfpoint{2\pgfplotmarksize}{.666667\pgfplotmarksize}}%
    \pgfpathrectangle
        {\pgfpoint{-.333333\pgfplotmarksize}{-\pgfplotmarksize}}
        {\pgfpoint{.666667\pgfplotmarksize}{2\pgfplotmarksize}}%
    \pgfusepathqfill
}

\newcommand{\SeriesMath}[2]{%
    \begingroup
    \pgfkeys{/pgf/fpu=true,/pgf/fpu/output format=fixed}%
    \pgfmathparse{#2}%
    \xdef#1{\pgfmathresult}%
    \endgroup
}

\newcommand{\PlotSeriesWithBox}[5]{%
    \gdef\SeriesCoords{}%
    \gdef\SeriesFound{0}%
    \pgfplotstableforeachcolumnelement{env}\of\latencydata
        \as\SeriesEnv{%
        \pgfplotstablegetelem{\pgfplotstablerow}{app}\of\latencydata
        \edef\SeriesApp{\pgfplotsretval}%
        \pgfplotstablegetelem{\pgfplotstablerow}{autoscaler}\of\latencydata
        \edef\SeriesAuto{\pgfplotsretval}%
        \ifdefstring{\SeriesEnv}{#1}{%
            \ifdefstring{\SeriesApp}{#2}{%
                \ifdefstring{\SeriesAuto}{#3}{%
                    \pgfplotstablegetelem
                        {\pgfplotstablerow}{replicas}\of\latencydata
                    \edef\SeriesX{\pgfplotsretval}%
                    \pgfplotstablegetelem
                        {\pgfplotstablerow}{median_p95_ms}\of\latencydata
                    \edef\SeriesY{\pgfplotsretval}%

                    \ifstrequal{#2}{TS}{%
                        \SeriesMath{\SeriesY}
                            {\SeriesY == 60000 ? 1200 : \SeriesY}%
                    }{}%

                    \xdef\SeriesCoords{%
                        \SeriesCoords(\SeriesX,\SeriesY)}%
                    \ifnum\SeriesFound=0
                        \xdef\SeriesXMin{\SeriesX}%
                        \xdef\SeriesXMax{\SeriesX}%
                        \xdef\SeriesYMin{\SeriesY}%
                        \xdef\SeriesYMax{\SeriesY}%
                        \gdef\SeriesFound{1}%
                    \else
                        \SeriesMath{\SeriesXMin}
                            {min(\SeriesXMin,\SeriesX)}%
                        \SeriesMath{\SeriesXMax}
                            {max(\SeriesXMax,\SeriesX)}%
                        \SeriesMath{\SeriesYMin}
                            {min(\SeriesYMin,\SeriesY)}%
                        \SeriesMath{\SeriesYMax}
                            {max(\SeriesYMax,\SeriesY)}%
                    \fi
                }{}%
            }{}%
        }{}%
    }%

    \ifnum\SeriesFound=1
        \SeriesMath{\BoxLeft}{max(0,\SeriesXMin-0.7)}%
        \SeriesMath{\BoxRight}{min(40,\SeriesXMax+0.7)}%
        \SeriesMath{\BoxBottom}{max(0,\SeriesYMin-40)}%
        \SeriesMath{\BoxTop}{min(1400,\SeriesYMax+40)}%

        \def\SeriesMarkSize{2.35pt}%
        \ifstrequal{#3}{HPA-Unfixed}{%
            \def\SeriesMarkSize{2.8pt}%
        }{}%

        \ifstrequal{#3}{WhatIf}{%
            \ifstrequal{#1}{dedicated}{%
                \def\SeriesMarkSize{3.2pt}%
            }{%
                \ifstrequal{#1}{shared}{%
                    \def\SeriesMarkSize{2.2pt}%
                }{}%
            }%
        }{}%

        \edef\RenderSeries{%
            \noexpand\addplot[
                no marks, draw=#4, fill=none,
                line width=0.8pt,
                dash pattern=on 4pt off 3.2pt,
                on layer=axis background, forget plot
            ] coordinates {
                (\BoxLeft,\BoxBottom)
                (\BoxRight,\BoxBottom)
                (\BoxRight,\BoxTop)
                (\BoxLeft,\BoxTop)
                (\BoxLeft,\BoxBottom)
            };
            \noexpand\addplot[
                only marks, mark=#5, mark size=\SeriesMarkSize,
                color=#4,
                mark options={draw=#4,fill=#4,line width=0.3pt},
                forget plot
            ] coordinates {\SeriesCoords};
        }%
        \RenderSeries
    \fi
}

\newcommand{\PlotApplication}[1]{%
    \PlotSeriesWithBox{dedicated}{#1}{Erlang}
        {dedicatedColor}{triangle*}%
    \PlotSeriesWithBox{dedicated}{#1}{HPA}
        {dedicatedColor}{*}%
    \PlotSeriesWithBox{shared}{#1}{Erlang}
        {sharedColor}{triangle*}%
    \PlotSeriesWithBox{shared}{#1}{HPA}
        {sharedColor}{*}%

    \PlotSeriesWithBox{dedicated}{#1}{WhatIf}
        {dedicatedColor}{filled plus}%
    \PlotSeriesWithBox{shared}{#1}{WhatIf}
        {sharedColor}{filled plus}%
}

\begin{figure}[!t]
\centering
\begin{tikzpicture}
\begin{groupplot}[
    group style={group size=1 by 3,vertical sep=1.45cm},
    scale only axis,
    width=0.8\columnwidth,
    height=3.175cm,
    xmin=5, xmax=40,
    ymin=0, ymax=1300,
    xtick={5,10,20,30,40,50},
    xticklabels={5,10,20,30,40,50},
    ytick={0,600,900,1200},
    yticklabels={0,600,900,1.2K},
    xlabel={Replicas},
    ylabel={Median p95 latency (ms)},
    label style={font=\small},
    tick label style={font=\footnotesize},
    tick align=inside,
    xtick pos=bottom,
    ytick pos=left,
    major tick length=3pt,
    tick style={draw=black!35,line width=0.6pt},
    axis line style={draw=black!35,line width=0.5pt},
    grid=major,
    grid style={draw=black!15,line width=0.5pt},
    extra y ticks={300},
    extra y tick labels={300},
    extra y tick style={
        grid=major,
        grid style={
            draw=red, line width=0.9pt,
            dash pattern=on 3.6pt off 2.7pt
        },
        tick label style={text=red}
    },
    set layers=standard,
    clip mode=individual,
    clip marker paths=true
]

\nextgroupplot[
    xlabel style={
        at={(axis description cs:0.5,0)},
        anchor=north,
        yshift=-15pt,       %
        inner sep=0pt
    }
]
\PlotApplication{OB}

\pgfplotsset{
    legend columns=3,
    legend cell align=left,
    legend style={
        at={(0.5,1.10)},
        anchor=south,
        font=\scriptsize,
        draw=black!25,
        fill=white,
        inner xsep=3pt,
        inner ysep=3pt,
        row sep=2pt,
        /tikz/every even column/.append style={
            column sep=3pt
        }
    }
}

\addlegendimage{
    line legend,
    draw=dedicatedColor,
    line width=1pt
}
\addlegendentry{Dedicated Host}

\addlegendimage{
    line legend,
    draw=white,
    line width=1pt
}
\addlegendentry{\phantom{Dedicated}}

\addlegendimage{
    line legend,
    draw=sharedColor,
    line width=1pt
}
\addlegendentry{Shared Host}

\addlegendimage{
    only marks,
    mark=*,
    mark size=2.7pt,
    mark options={solid,draw=black,fill=black}
}
\addlegendentry{HPA}

\addlegendimage{
    only marks,
    mark=triangle*,
    mark size=2.7pt,
    mark options={solid,draw=black,fill=black}
}
\addlegendentry{Erlang}

\addlegendimage{
    only marks,
    mark=filled plus,
    mark size=2.7pt,
    mark options={solid,draw=black,fill=black}
}
\addlegendentry{\name}

\nextgroupplot[
    xlabel style={
        at={(axis description cs:0.5,0)},
        anchor=north,
        yshift=-15pt,       %
        inner sep=0pt
    }
]
\PlotApplication{SN}

\nextgroupplot[
    yticklabels={0,600,900,{60K}}
]
\PlotApplication{TS}

\end{groupplot}

\coordinate (tsbreak) at
    ($(group c1r3.south west)!0.81!
      (group c1r3.north west)$);

\draw[
    overlay,
    white,
    line width=3pt
]
    ($(tsbreak)+(0,-4pt)$) --
    ($(tsbreak)+(0,4pt)$);

\draw[
    overlay,
    black,
    line width=0.7pt,
    line cap=round,
    line join=round
]
    ($(tsbreak)+(0,-4pt)$) --
    ($(tsbreak)+(-2pt,-2pt)$) --
    ($(tsbreak)+(2pt,0pt)$) --
    ($(tsbreak)+(-2pt,2pt)$) --
    ($(tsbreak)+(0,4pt)$);

\node[
    font=\small\bfseries,
    anchor=south,
    inner sep=0pt,
    outer sep=0pt
]
    at ([yshift=6pt]group c1r2.north)
    {(a) Online Boutique};

\node[
    font=\small\bfseries,
    anchor=south,
    inner sep=0pt,
    outer sep=0pt
]
    at ([yshift=6pt]group c1r3.north)
    {(b) Social Network};

\node[font=\small\bfseries]
    at ([yshift=-35pt]group c1r3.south)
    {(c) TeaStore};
    
\end{tikzpicture}

\caption{Per-endpoint median p95 latency when using HPA, Erlang, and \name on AWS EC2 \texttt{t3.2xlarge} instances with Dedicated and Shared Host tenancy. The horizontal dashed line marks the 300\,ms \textcolor{red!70!black}{SLO}. For each application, the boundary boxes mark the average replica count for each autoscaler–environment pair.}
\label{fig:autoscaler-latencies-aws}

\end{figure}
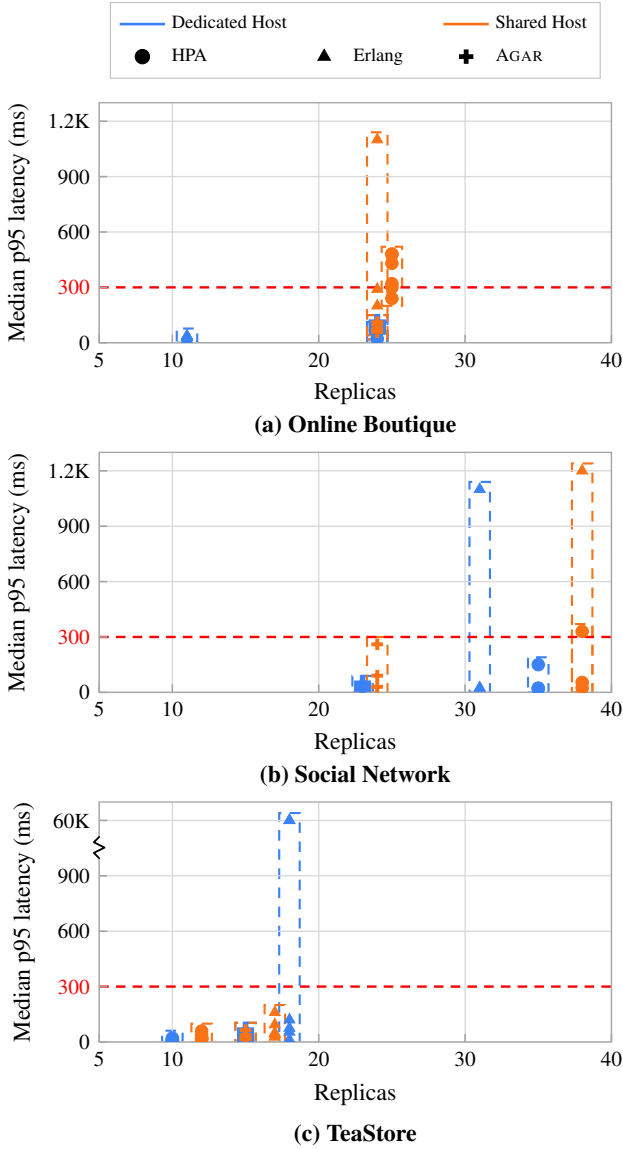

We set a 300\,ms p95 latency SLO and consider a violation when \emph{any} endpoint's p95 latency exceeds this target. Erlang instead uses a single p95 latency computed over requests across all endpoints as its latency signal, so an endpoint can violate the 300\,ms SLO without necessarily causing Erlang's measured p95 to exceed 300\,ms, as we show below. HPA is not SLO-aware and reacts directly to resource-utilization targets; consequently, satisfying its utilization targets does not imply SLO compliance.

\paragraph{Online Boutique.} 
\name keeps per-endpoint p95 latencies below 50\,ms on dedicated hosts and 100\,ms on shared hosts, well below the 300\,ms SLO. In contrast, Erlang can miss severe latency violations at infrequently requested endpoints. On shared hosts, for example, an endpoint that constitutes less than 0.6\% of requests reaches a median p95 latency of $\sim$1.1\,s and violates the SLO, even though Erlang’s overall p95 latency remains below it. Because Erlang computes this percentile across all endpoints rather than separately for each endpoint, sufficiently rare slow endpoints can be masked by the larger volume of lower-latency requests. Erlang cannot distinguish between endpoints, and any workload where fast endpoints constitute more than $1-p$ of traffic—where $p$ is the tail latency percentile—will expose this limitation again. Figure~\ref{fig:autoscaler-latencies-aws} reports the endpoint-level latency and corresponding average replica cost for each autoscaler.

\paragraph{TeaStore.}
With Erlang, the same problem manifests in TeaStore, though in a more extreme form. Two endpoints account for $<$0.9\% and $<$0.4\% of total traffic. Requests for these endpoints time out, violating the 300\,ms SLO. However, at under 1.3\% of total traffic, their requests fall above the 95th percentile of the aggregate latency distribution—well out of reach of Erlang’s p95 reward signal. The aggregate p95 is determined entirely by four other endpoints that collectively account for over 98\% of traffic—and registers no violation. Erlang therefore has no incentive to provision for the violating endpoints and converges to a policy that satisfies the aggregate target while leaving them in violation. The difference is one of degree: in Online Boutique, latency violations at infrequently requested endpoints are masked by the larger volume of lower-latency requests, whereas in TeaStore the violating endpoints are so rare that they lie entirely beyond the percentile Erlang optimizes for. 

\pgfplotstableread[col sep=comma]{data/SN-workload-change.csv}\snworkload

\definecolor{Eone}{HTML}{0072B2}
\definecolor{Etwo}{HTML}{D55E00}
\definecolor{Ethree}{HTML}{009E73}
\definecolor{P95}{HTML}{E85BC0}

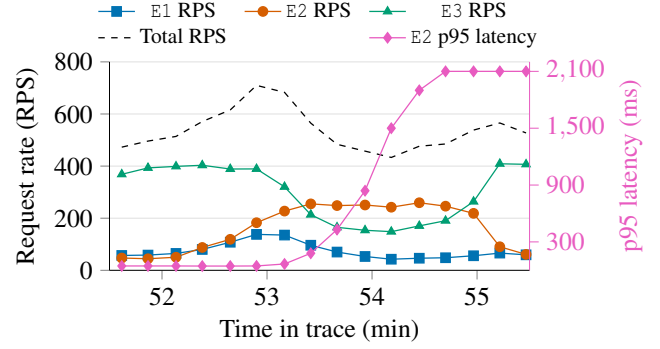
\begin{figure}[!t]
\noindent\makebox[\columnwidth][c]{%
\resizebox{\columnwidth}{!}{%
\begin{tikzpicture}[font=\large]

\begin{axis}[
  width=\columnwidth,height=5cm,
  xmin=51.5,xmax=55.5,ymin=0,ymax=800,
  xtick={52,53,54,55},ytick={0,200,400,600,800},
  xlabel={Time in trace (min)},
  ylabel={Request rate (RPS)},
  axis lines=left,
  axis line style={-},
  legend style={
    at={(0.5,1.03)},anchor=south,
    draw=none,font=\normalsize,
    /tikz/every even column/.append style={column sep=6pt}
  },
  ymajorgrids,grid style={gray!25},
  tick align=outside,
  every axis plot/.append style={semithick,mark size=2.2pt},
  legend columns=3,
]

\addplot[Eone,mark=square*]
  table[x=minutes,y=E1_rps]{\snworkload};
\addlegendentry{\texttt{E1} RPS}

\addplot[Etwo,mark=*]
  table[x=minutes,y=E2_rps]{\snworkload};
\addlegendentry{\texttt{E2} RPS}

\addplot[Ethree,mark=triangle*]
  table[x=minutes,y=E3_rps]{\snworkload};
\addlegendentry{\texttt{E3} RPS}

\addplot[black,dashed,mark=none]
  table[x=minutes,y=total_rps]{\snworkload};
\addlegendentry{Total RPS}

\addlegendimage{empty legend}
\addlegendentry{}
\addlegendimage{P95,semithick,mark=diamond*,mark size=2.4pt}
\addlegendentry{\texttt{E2} p95 latency}

\end{axis}

\begin{axis}[
  width=\columnwidth,height=5cm,
  xmin=51.5,xmax=55.5,
  ymin=0,ymax=2200,
  ytick={300,900,1500,2100},
  axis y line*=right,
  axis x line=none,
  ylabel={p95 latency (ms)},
  axis line style={P95},
  ytick style={P95},
  yticklabel style={text=P95},
  ylabel style={text=P95},
  tick align=outside,
  every axis plot/.append style={semithick,mark size=2.4pt},
]

\addplot[P95,mark=diamond*]
  table[x=minutes,y=p95_ms]{\snworkload};

\end{axis}

\end{tikzpicture}%
}
}

\caption{Workload transition for Social Network: \texttt{E2}'s request rate rises as \texttt{E3}'s and the aggregate request rate decline. Since Erlang scales based on aggregate request rate, \texttt{E2}'s p95 latency exceeds the 300\,ms SLO.}
\label{fig:sn-workload-change}
\end{figure}

\paragraph{Social Network.} 

Around 53 minutes into the run, endpoint \texttt{E2}'s request rate rises from roughly 40 to 250 RPS, while \texttt{E3}'s falls from roughly 400 to 150 RPS and the aggregate request rate declines, as shown in Figure~\ref{fig:sn-workload-change}. During this period, \name adjusts its replica allocation
as demand shifts between endpoints. Erlang's allocation remains unchanged due to reduction in total RPS: its scaling trigger uses aggregate RPS and ignores endpoint mix and latency. \texttt{E2}'s recorded cumulative p95 rises above the 300\,ms SLO and stays above it until the end of the run, yielding median values of 1.1\,s on dedicated hosts and 1.2\,s on shared hosts.

\subsection{\name as a Diagnostic Tool}\label{subsec:debugger}

While AWS serves as our primary environment for evaluating autoscaling, the applications exhibited diverse performance problems on a CloudLab~\cite{cloudlab} cluster of r320 machines. The Online Boutique (\S\ref{case-study-ob}) and TeaStore (\S\ref{case-study-ts}) issues did not occur on AWS, which we attribute to its substantially faster storage; the Social Network issue (\S\ref{case-study-sn}) occurred in both environments. Thus, the case studies span issues that arise from interactions with the underlying infrastructure as well as those that persist across environments that \name was able to diagnose.

\subsubsection{Online Boutique}\label{case-study-ob}

We begin this case study with a baseline comparison: we first observe how standard autoscalers behave under the workload, then use \name to perform counterfactual diagnosis and isolate the structural cause.

When we ran Alibaba workloads for Online Boutique, SLO violations were observed for two endpoints. Both HPA and Erlang responded to these violations by scaling out: HPA increased replica counts for services exceeding their CPU thresholds, while Erlang adjusted replica counts across services according to its learned autoscaling policy. 

Despite this, both approaches failed to alleviate SLO violations: as load increased, newly scheduled replicas could not serve traffic until their initialization completed, concentrating load on already-running replicas and causing queue buildup.

\name took a different approach. Since no observed configuration satisfied the SLO, we reasoned counterfactually by fitting the utilization models assuming $T_{i}=100.0$ for every service $i$ (see~\S\ref{subsec:joint-optimization}) — the most optimistic possible scaling threshold to meet an SLO. Even under this assumption, \texttt{recommendation}\allowbreak\texttt{service}, \texttt{payment}\allowbreak\texttt{service}, and \texttt{currency}\allowbreak\texttt{service} exhibited anomalously high $\beta$ values -- high CPU utilization kept triggering scale-outs, but newly scheduled replicas that failed their probes served no requests. This directed our attention to these services' configurations rather than their replica counts.

The root cause was a probe timeout misconfiguration. Although the same configuration worked on AWS, CloudLab’s substantially slower storage—roughly 153 IOPS compared with 1636 IOPS on AWS—increased service initialization times enough for the probes to time out before replicas became ready.

Kubernetes readiness probes~\cite{k8s_configure_probes} with insufficient timeouts prevented replicas from receiving traffic, concentrating load on the replicas that do pass the probes. Since load was concentrated on fewer replicas than were provisioned, the model underestimated how much CPU utilization was due to request load, and the remainder appeared as inflated $\beta$. Liveness probe failures compounded this by triggering \texttt{CrashLoopBackOff} restarts, further destabilizing the pool. Since each new replica was susceptible to probe failure, the SLO was not violated when the HPA was configured with \texttt{minReplicas: 10}. Yet overprovisioned resources did not solve the underlying problem. Increasing the probe timeout to match the service’s actual startup latency immediately stabilized the pool, restored normal load distribution, and satisfied the SLO with the minimum replica count prescribed by \name.

This example shows that autoscaling was not the solution. The violation was not fundamentally a capacity problem, but an environment-dependent configuration problem: probe timeouts that were sufficient on AWS became too short on CloudLab, where slower storage increased service startup times. \name helped localize the root cause and avoid unnecessary scaling.

\vspace{-2mm}

\subsubsection{TeaStore}\label{case-study-ts}

When we ran Alibaba workloads for TeaStore, SLO violations were observed across all endpoints. HPA continued scaling out \texttt{teastore-webui}, TeaStore's client-facing frontend, in response to its high CPU utilization. The SLO was eventually satisfied after a prolonged queue buildup period and at significant replica cost. From HPA's perspective, the system was simply underprovisioned.

Since the SLO was eventually satisfied after the queue buildup released, our SLO model produced SLO-safe scaling thresholds for all TeaStore services. When we fit the utilization models (as described in~\S\ref{subsec:joint-optimization}) with these thresholds and the corresponding observed workload, \texttt{teastore-webui} emerged as the sole outlier — all other services exhibited normal $\beta$ and $\epsilon$ values, while \texttt{teastore-webui} showed anomalously high $\epsilon$ on both CPU and memory utilization. 

The root cause was an unbounded JVM heap. Without explicit heap limits, the JVM expanded the heap and metaspace to the container’s memory limit, causing garbage collection (GC) cycles to grow unpredictably and consume substantial CPU independent of request load. This configuration did not violate the SLO on AWS, but CloudLab’s substantially slower storage amplified the resulting GC overhead and prolonged queue buildup, exposing the problem. The resulting utilization variability manifested as anomalously high $\epsilon$ in the utilization model. Unlike the probe misconfiguration—where load concentration on fewer replicas artificially inflated $\beta$—here the erratic GC cycles directly drove utilization variability. Bounding the heap to three-quarters of the container’s memory limit and metaspace to one-eighth via \texttt{JAVA\_TOOL\_OPTIONS}~\cite{oracle-jvm-tuning} ensured predictable GC behavior, making CPU utilization a function of application load rather than JVM ergonomics.

\subsubsection{Social Network}\label{case-study-sn}

When we ran Alibaba workloads for Social Network with HPA, the SLO was satisfied throughout — from the perspective of any existing tool, the system was healthy. Yet when \name fit its models on the collected data, \texttt{user-timeline-redis} exhibited anomalously high $\epsilon$ on memory utilization. Unlike the Online Boutique and TeaStore issues, this anomaly appeared on both CloudLab and AWS, showing that it was not confined to the slower CloudLab environment.

\begin{figure}[H]
    \centering
    \includegraphics[width=\columnwidth]{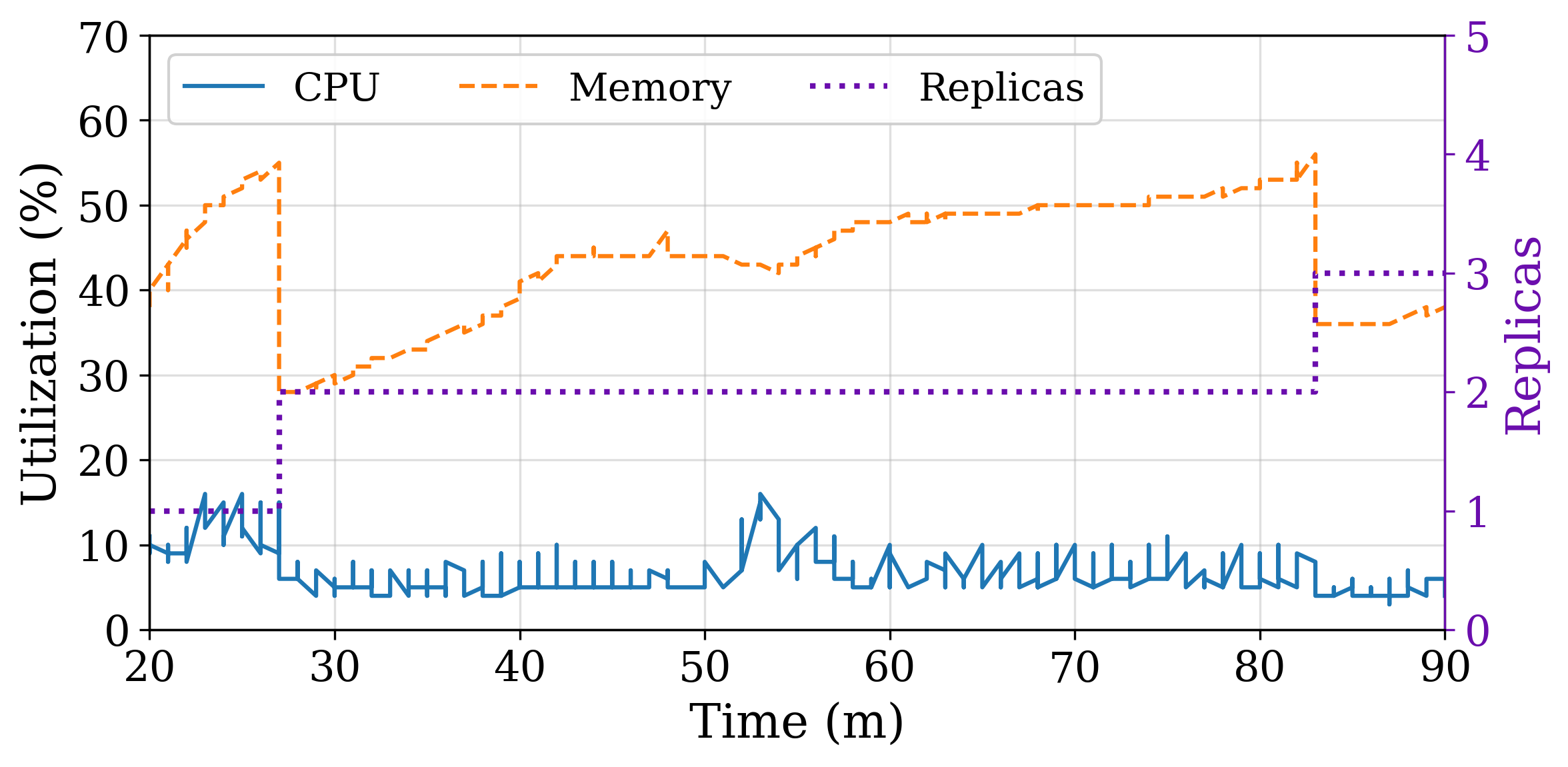}
    \vspace{-5mm}
    \caption{Memory utilization and replica count for \texttt{user-timeline-redis}. Memory grows monotonically with data, triggering repeated HPA scale-outs that provide only temporary relief before memory resumes climbing.}
    \label{fig:redis-utilization}    
\end{figure}

The anomaly is explained by the memory utilization pattern visible in Figure~\ref{fig:redis-utilization}: memory climbs steadily, punctuated by sharp drops each time HPA scales out and dilutes the per-pod average, before resuming its climb.

The root cause is a misconfiguration chain: \texttt{redis.conf} sets no \texttt{maxmemory} bound, defaulting to Redis's \texttt{noeviction} policy, so data accumulates with every write and memory grows monotonically. HPA, configured with a memory scaling threshold, mistakes this growth for load and scales the service out — providing only temporary relief as the dilution effect fades and memory resumes climbing, driving further scale-outs indefinitely. Notably, when HPA is configured to scale on CPU rather than memory with the same \texttt{redis.conf}, Redis accumulates data until the container's memory limit is exhausted and the pod is OOMKilled. 

Setting \texttt{maxmemory 100mb} in \texttt{redis.conf} caps memory usage within the container's 128Mi limit, and \texttt{maxmemory\-policy} \texttt{allkeys-lru} ensures Redis evicts the least recently used entries when the cap is reached rather than accumulating without bound — the specific policy matters less than having one. With these set, memory stabilizes once the working set fills the cap, eliminating the scale-out loop at its root.

\vspace{-2mm}
\subsection{Other Counterfactual Questions}

\subsubsection{SLO--cost trade-off.}
\label{s:slo-cost}

\begin{figure}[!t]
\centering

\definecolor{sharedColor}{HTML}{3B82F6}
\definecolor{dedicatedColor}{HTML}{F97316}

\pgfplotstableread[col sep=comma]
    {data/slo-sweep-OB-AWS.csv}\slodata

\pgfplotstableforeachcolumnelement{env}\of\slodata\as\envvalue{%
    \pgfplotstablegetelem{\pgfplotstablerow}{slo}\of\slodata
    \edef\slovalue{\pgfplotsretval}%

    \ifnum\slovalue=200\relax
        \ifdefstring{\envvalue}{dedicated}{%
            \pgfplotstablegetelem{\pgfplotstablerow}{sim_cost}\of\slodata
            \xdef\DedicatedSim{\pgfplotsretval}%
            \pgfplotstablegetelem{\pgfplotstablerow}{real_cost}\of\slodata
            \xdef\DedicatedReal{\pgfplotsretval}%
        }{}%

        \ifdefstring{\envvalue}{shared}{%
            \pgfplotstablegetelem{\pgfplotstablerow}{sim_cost}\of\slodata
            \xdef\SharedSim{\pgfplotsretval}%
            \pgfplotstablegetelem{\pgfplotstablerow}{real_cost}\of\slodata
            \xdef\SharedReal{\pgfplotsretval}%
        }{}%
    \fi
}

\begin{tikzpicture}
\begin{axis}[
    width=0.8\columnwidth,
    scale only axis=true,
    height=4.3cm,
    ylabel={Cost (vCPUs/hr)},
    ylabel style={font=\small, yshift=-5pt},
    ymin=0,
    ymax=35,
    ytick={0,5,10,15,20,25,30,35},
    yticklabel style={font=\small},
    symbolic x coords={Dedicated,Shared},
    xtick={Dedicated,Shared},
    xticklabels={Dedicated,Shared},
    xticklabel style={font=\small},
    axis y line*=left,
    axis x line*=bottom,
    axis line style={gray!50},
    tick style={gray!50},
    grid=major,
    xmajorgrids=false,
    grid style={gray!15, thin},
    bar width=15pt,
    enlarge x limits=0.35,
    legend style={
        at={(0.5,1.02)},
        anchor=south,
        font=\small,
        draw=gray!30,
        fill=white,
        legend columns=2,
        column sep=10pt,
    },
]

\addplot[
    ybar,
    bar shift=-8pt,
    fill=white,
    draw=dedicatedColor,
    thick,
    pattern=north east lines,
    pattern color=dedicatedColor,
    forget plot,
]
coordinates {(Dedicated,\DedicatedSim)};

\addplot[
    ybar,
    bar shift=8pt,
    fill=dedicatedColor,
    draw=dedicatedColor,
    thick,
    forget plot,
]
coordinates {(Dedicated,\DedicatedReal)};

\addplot[
    ybar,
    bar shift=-8pt,
    fill=white,
    draw=sharedColor,
    thick,
    pattern=north east lines,
    pattern color=sharedColor,
    forget plot,
]
coordinates {(Shared,\SharedSim)};

\addplot[
    ybar,
    bar shift=8pt,
    fill=sharedColor,
    draw=sharedColor,
    thick,
    forget plot,
]
coordinates {(Shared,\SharedReal)};

\addlegendimage{
    area legend,
    draw=black,
    fill=white,
    pattern=north east lines,
    pattern color=black
}
\addlegendentry{Simulated}

\addlegendimage{
    area legend,
    draw=black,
    fill=black
}
\addlegendentry{Real}

\end{axis}
\end{tikzpicture}

\caption{Cost to meet a 200\,ms p95 latency target for Online Boutique across dedicated and shared environments. Cost is based on the hourly number of vCPUs required to support \name's prescribed replica configuration. Real costs are higher due to online retraining.}
\label{fig:slo-cost}
\end{figure}
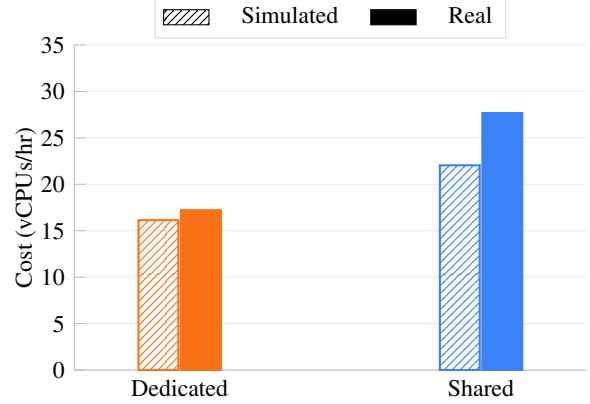

Equation~\ref{eqn:num-pods} makes the SLO target $\tau$ an explicit parameter, so \name can answer: \emph{given this workload, what SLO can I afford?} To evaluate this, we sweep $\tau$ over a range of p95 latency targets and compute the hourly vCPU cost by replaying the observed RPS from an unseen trace through Equation~\ref{eqn:num-pods} at each tick---without re-running the cluster. We validate these simulated estimates against live autoscaler runs on the same workload at these SLO targets. Figure~\ref{fig:slo-cost} compares predicted and measured costs to meet a 200\,ms SLO for Online Boutique across the dedicated and shared environments. Using models trained separately on data from each environment, \name predicts that meeting an SLO of 200\,ms requires 36.6\% more vCPUs on \emph{t3.2xlarge} instances with shared tenancy than Dedicated Host tenancy.

\subsubsection{Varying a Service's Pod Startup Delay}

The pod startup delay $\Delta C_j$ (see~\S\ref{sec:modeling-philosophy}) is used to generate the numerator of Equation~\ref{eqn:num-pods}: a longer startup window requires the autoscaler to provision for a higher peak RPS, since new pods cannot serve traffic until initialization completes. \name can therefore answer the counterfactual: \emph{how does reducing a service's startup time affect cost?}

For TeaStore, this effect is specific to the service \texttt{teastore}\allowbreak\texttt{-webui}. Reducing its startup time from 30\,s to 15\,s lowers the cost of meeting a 300\,ms SLO from 12 to 8 vCPUs/hr on \texttt{t3.2xlarge} instances with Dedicated Host tenancy and from 14 to 10 vCPUs/hr on instances with shared tenancy, corresponding to 33\% and 29\% reductions, respectively. The reduction comes from shrinking the RPS window over which the autoscaler must provision replicas. Applying the same optimization to other services yields no cost benefit, allowing developers to prioritize optimization effort.

\vspace{-2mm}

\subsection{Training Cost}\label{eval:training-cost}

Training time has two components. The first is generating candidate threshold configurations for the SLO satisfaction model in a threshold space of dimension $d$ = (number of services) $\times$ (number of resource types). Although threshold search can in principle become expensive with  $d$, threshold generation is not the bottleneck in our implementation. This stage performs dominance filtering over observed SLO-violating utilization points followed by a cost-aware greedy search. The dominant cost arises when \name trains utilization models requested by this search. These fits are performed on demand, potentially in multiple batches at the same SLO, and cached by service, resource type, and threshold across the SLO sweep.

For each threshold configuration, \name solves two LPs per service-resource pair: the first minimizes the transformed cost objective (Equation~\ref{eqn:num-pods}), and the second fixes that optimum and minimizes $\epsilon_{i}$. The size of each LP is governed primarily by the number of endpoints and the number of training samples ($H$), since the formulation introduces one $\alpha_{i,j}$ variable for each timestep-endpoint pair, for a total of $\#\text{endpoints}\times H$ such variables, together with the associated monotonicity constraints. In our implementation, each CBC solve is single-threaded, and parallelism comes from the outer \texttt{joblib}~\cite{joblib} layer across independent model fits, using two workers for batches and direct execution for singleton fits. 

Retraining can either be: (1) full retraining, which repeats the training procedure above, or (2) utilization-bound retraining, which refits only the affected utilization model (described in~\S\ref{sec:modeling-philosophy}).
\begin{table}[!t]
\centering
\small
\setlength{\tabcolsep}{3pt}
\begin{tabularx}{\columnwidth}{
  @{}|c|c|c|
  >{\centering\arraybackslash}X|
  >{\centering\arraybackslash}X|@{}
}
\hline
\multirow[c]{3}{*}{\textbf{Env}} &
\multirow[c]{3}{*}{\textbf{App}} &
\multirow[c]{3}{*}{$\mathbf{H}$} &
\multicolumn{2}{c|}{\textbf{Training time (s)}} \\
\cline{4-5}
& & &
\shortstack{\rule{0pt}{2.1ex}\textbf{Full}\\\textbf{retraining}} &
\shortstack{\rule{0pt}{2.1ex}\textbf{Utilization bound}\\\textbf{retraining}} \\
\hline
\multirow[c]{6}{*}{Shared}
& \multirow[c]{2}{*}{SN}
& 500 & 65.22 & 1.16--1.88 \\
\cline{3-5}
& & 1000 & 235.84 & 1.58--5.56 \\
\cline{2-5}
& \multirow[c]{2}{*}{TS}
& 500 & 41.60 & 0.67--1.26 \\
\cline{3-5}
& & 1000 & 125.19 & 1.35--2.53 \\
\cline{2-5}
& \multirow[c]{2}{*}{OB}
& 500 & 293.23 & 1.33--4.35 \\
\cline{3-5}
& & 1000 & 811.23 & 2.64--14.95 \\
\hline
\multirow[c]{6}{*}{Dedicated}
& \multirow[c]{2}{*}{SN}
& 500 & 56.18 & 0.89--1.33 \\
\cline{3-5}
& & 1000 & 167.19 & 1.21--3.84 \\
\cline{2-5}
& \multirow[c]{2}{*}{TS}
& 500 & 25.27 & 0.48--1.51 \\
\cline{3-5}
& & 1000 & 78.15 & 0.89--2.23 \\
\cline{2-5}
& \multirow[c]{2}{*}{OB}
& 500 & 73.69 & 0.95--1.21 \\
\cline{3-5}
& & 1000 & 222.61 & 1.84--11.52 \\
\hline
\end{tabularx}
\caption{Full retraining, which is equivalent to initial training, reports the median of three runs, while utilization-bound retraining reports the range of median latencies across utilization models. Shared and Dedicated denote EC2 tenancy, and SN, TS, and OB denote Social Network, TeaStore, and Online Boutique, respectively. $H$ is the number of training samples.}\label{tab:training-time}
\end{table}

Table~\ref{tab:training-time} summarizes training and retraining costs. Across all applications and environments, median full-training
latency is at most 293.23\,s at $H=500$ and 811.23\,s at $H=1000$, while median utilization-bound retraining latency is at most 14.95\,s, just below the \name autoscaler's 15\,s decision interval.

\section{Related Work}

\noindent\textbf{Counterfactual performance modeling and what-if analysis.}
A related line of work uses causal or trace-driven models to answer hypothetical performance questions in complex systems. CausalSim~\cite{causalsim} and Veritas~\cite{veritas} develop counterfactual reasoning methods for simulating networks. CausalSim learns system dynamics from randomized experiments to correct for the fact that an intervention can itself change the trace being replayed. Veritas in contrast avoids randomized experiments by constructing a domain-specific causal model that captures latent network conditions, trading experimental effort for modeling effort. Thesios~\cite{thesios} synthesizes representative and counterfactual full-resolution I/O traces from down-sampled storage traces, enabling what-if analyses of hypothetical policy and hardware changes. These systems share our interest in reasoning about unobserved scenarios, but they target different systems and do not compose per-component models. Our setting requires composition of several models. %
WebPerf~\cite{webperf} predicts request-latency distributions under hypothetical cloud configurations and workloads by combining request-dependency graphs with profiled cloud-API latencies, while \name reasons about SLO feasibility using per-service bounds learned from application measurements. Slowpoke~\cite{slowpoke} evaluates how optimizing a service affects application \emph{throughput} by actively slowing down every other service to create an application with similar bottleneck structure, and avoids having to answer counterfactual questions from only passively collected data. Further, it answers a different question than \name. It focuses on a service's throughput while we reason about its startup latency, among other things. TierSizer~\cite{tiersizer} uses analytical models to predict application slowdown at unexecuted DRAM sizes, enabling counterfactual DRAM sizing without instantiating those configurations, but like other work, this setting does not need models to compose. %

\noindent\textbf{Autoscaling and resource management.}
A large body of work studies SLO-aware autoscaling and resource allocation for microservices. Most of it focuses on vertical rather than horizontal autoscaling; \name's autoscaler targets the latter. These systems scale CPU and memory resources within a VM, rather than the number of VMs. Further, \name is more than an autoscaler. It can determine whether an SLO can be guaranteed under a given workload, what provisioning that guarantee requires, what it costs, and when scaling cannot help because the bottleneck is structural.

Systems such as Sinan~\cite{sinan-delimitrou} and FIRM~\cite{firm-osdi} use telemetry-driven machine learning to identify bottlenecks and \emph{vertically} adjust per-service resources to maintain end-to-end latency targets. FIRM primarily performs vertical scaling, using scale-out only under resource oversubscription. AutoThrottle~\cite{autothrottle} similarly uses \emph{vertical} scaling to enable online adaptation of per-service resources to application-level SLOs. It avoids the combinatorial cost of jointly optimizing per-service resources using a bi-level controller, where local controllers translate application-level throttle ratios into CPU allocations, at the cost of potentially missing more resource-efficient allocations outside its restricted search space. Erlang~\cite{erlang} jointly allocates resources across all microservices of an application to minimize cost while satisfying an end-to-end latency target. GRAF~\cite{graf} similarly targets end-to-end latency, using a graph neural network to proactively allocate resources based on workload changes. More structured approaches such as Ursa~\cite{ursa} and Erms~\cite{erms} use analytic or piecewise-linear models to map per-service behavior to resource allocations and SLO satisfaction. Ursa actively searches for per-service utilization thresholds by reducing resources under replayed workloads until latency degrades, while Erms profiles piecewise-linear latency models over workload, CPU allocation, and interference for individual microservices and their dependencies. Galileo~\cite{percs} uses a queueing-theoretic model to derive performance robustness certificates for request-type tail latencies under bounded environmental perturbations, and uses these certificates to guide learned autoscaling and admission control. These systems are primarily designed to choose or optimize allocations online. 

\noindent\textbf{Debugging and root-cause analysis.} Distributed tracing systems like Jaeger~\cite{jaeger_uber_2017} and Zipkin~\cite{zipkin_twitter_2012} can be used to, among other things, identify the service which is experiencing high latency. To mitigate the instrumentation cost of tracing, prior work has explored tomography~\cite{traceweaver}. Murphy~\cite{murphy} uses a Markov random field to identify candidate root causes of performance problems from monitored metrics and provides interpretable explanations. Sage~\cite{sage} identifies the service which caused an SLO violation and fixes it by reallocating resources (including compute, memory, cache or network resources) or reducing contention. However, it only focuses on problems that can be fixed by adding more resources to the problem. By developing an interpretable and global model of the application, \name can identify situations where an application is structurally incapable of reliably satisfying an SLO through autoscaling alone. Another line of work has focused on \emph{negative provenance} to diagnose correctness (as opposed to performance) issues in networked systems~\cite{negative-provenance, diffprov}.

\noindent\textbf{Formal modeling and verification.}
Another line of work reasons about microservice applications through explicit formal models. Kivi~\cite{kivi} model-checks Kubernetes controllers and configurations by exploring controller/event interleavings. Anvil~\cite{anvil} verifies that controller implementations satisfy \emph{eventually stable reconciliation}, a liveness property. Both target correctness, rather than performance properties. Performal~\cite{performal} is closer in spirit to \name than prior verification work: like \name, it uses upper bounds and can surface performance bugs. However, its bounds concern the worst-case end-to-end latency of single-request distributed systems, which is proved symbolically from a formal model of the distributed algorithm and then related to real time using isolated component measurements. Recently, a line of work on performance verification has used assumption-constrained worst-case models to find pathological behaviors in existing algorithms~\cite{ccac, metaopt, xplain, virelay} and synthesize new ones~\cite{syntra, ccmatic, worst-case-llm-heuristics}. Relatedly, another line of work analyzes program executions to determine worst-case running time and memory access patterns~\cite{perf-contract-nf, adversarial-nf}. All of these use formal reasoning over detailed models of the algorithms used and are hence limited in scope. \name is more data-driven and can thus reason about end-to-end performance of a microservice application without requiring detailed models of its implementation.

\vspace{-4mm}

\section{Discussion and Limitations}

\name assumes that the fundamental behavior of the application remains stable over time. In deployment, we therefore recommend retraining the models after each code change and whenever the underlying hardware changes.

In settings with substantial measurement noise or outliers, operators may choose to introduce slack in the models. Our framework can incorporate slack in all three models, allowing a small fraction of data points to violate the learned bounds. In our evaluation, slack was unnecessary for the utilization and SLO satisfaction models, as the applications were well-behaved. However, if this assumption does not hold, slack can be introduced with minimal changes, as our implementation already supports it.

Future work could extend \name to support vertical autoscaling and dynamic resource reallocation. More broadly, \name's approach could apply to other systems problems where component-level relationships can be learned and composed to answer end-to-end counterfactual questions. Supporting richer models will require better abstractions and training methods for specifying and learning worst-case system behavior. LLMs could assist in constructing such models, but the resulting relationships must still provide explicit bounds that can be validated and composed. While composing \name models to reason about end-to-end guarantees was straightforward in our setting, more complex systems may require automated support. Developing general-purpose languages and algorithms for specifying and training such models is an important direction for future work.

\vspace{-4mm}

\section{Conclusion}
We presented \name, a counterfactual performance analysis tool for microservice applications. \name builds simple models with explicit uncertainty bounds from readily available telemetry to reason about SLO feasibility under hypothetical changes to workloads, resource allocations, and SLO targets. We demonstrate how autoscaling and performance diagnosis can both be formulated as counterfactual queries that provide actionable guidance to operators.

\FloatBarrier
\bibliographystyle{plain}
\bibliography{refs}

\begin{thebibliography}{10}

\bibitem{ccmatic}
Anup Agarwal, Venkat Arun, Devdeep Ray, Ruben Martins, and Srinivasan Seshan.
\newblock Towards provably performant congestion control.
\newblock In {\em USENIX NSDI}, 2024.

\bibitem{causalsim}
Abdullah Alomar, Pouya Hamadanian, Arash Nasr-Esfahany, Anish Agarwal, Mohammad
  Alizadeh, and Devavrat Shah.
\newblock {CausalSim}: A causal framework for unbiased {Trace-Driven}
  simulation.
\newblock In {\em 20th USENIX Symposium on Networked Systems Design and
  Implementation (NSDI 23)}, pages 1115--1147, Boston, MA, April 2023. USENIX
  Association.

\bibitem{aws-dedicated-hosts}
{Amazon Web Services}.
\newblock {Amazon EC2 Dedicated Hosts}.
\newblock \url{https://aws.amazon.com/ec2/dedicated-hosts/}.
\newblock Accessed: 2026-09-01.

\bibitem{aws-nitro-side-channels}
{Amazon Web Services}.
\newblock The {EC2} approach to preventing side-channels.
\newblock
  \url{https://docs.aws.amazon.com/whitepapers/latest/security-design-of-aws-nitro-system/the-ec2-approach-to-preventing-side-channels.html}.

\bibitem{aws-burstable-credits}
{Amazon Web Services}.
\newblock Key concepts for burstable performance instances.
\newblock
  {\url{https://docs.aws.amazon.com/AWSEC2/latest/UserGuide/burstable-credits-baseline-concepts.html}}.

\bibitem{aws-autoscaling}
{Amazon Web Services}.
\newblock Target tracking scaling policies for {Amazon} {EC2} auto scaling.
\newblock
  \url{https://docs.aws.amazon.com/autoscaling/ec2/userguide/as-scaling-target-tracking.html},
  2024.
\newblock Accessed: 2025-06-20.

\bibitem{ccac}
Venkat Arun, Mina~T. Arashloo, Ahmed Saeed, Mohammad Alizadeh, and Hari
  Balakrishnan.
\newblock Toward formally verifying congestion control behavior.
\newblock In {\em ACM SIGCOMM}, 2021.

\bibitem{traceweaver}
Sachin Ashok, Vipul Harsh, Brighten Godfrey, Radhika Mittal, Srinivasan
  Parthasarathy, and Larisa Shwartz.
\newblock Traceweaver: Distributed request tracing for microservices without
  application modification.
\newblock In {\em Proceedings of the ACM SIGCOMM 2024 Conference}, pages
  828--842, 2024.

\bibitem{uber-failover-architecture}
Mayank Bansal, Milind Chabbi, Kenneth B{\o}gh, Srikanth Prodduturi, Kevin Xu,
  Amit Kumar, David Bell, Ranjib Dey, Yufei Ren, Sachin Sharma, Juan Marcano,
  Shriniket Kale, Subhav Pradhan, Ivan Beschastnikh, Miguel Covarrubias,
  Chien-Chih Liao, Sandeep~Koushik Sheshadri, Wen Luo, Kai Song, Ashish Samant,
  Sahil Rihan, Nimish Sheth, Albert Greenberg, and Uday~Kiran Medisetty.
\newblock Uber{\textquoteright}s failover architecture: Reconciling reliability
  and efficiency in hyperscale microservice infrastructure.
\newblock In {\em 23rd USENIX Symposium on Networked Systems Design and
  Implementation (NSDI 26)}, pages 2011--2026, Renton, WA, May 2026. USENIX
  Association.

\bibitem{cilantro}
Romil Bhardwaj, Kirthevasan Kandasamy, Asim Biswal, Wenshuo Guo, Benjamin
  Hindman, Joseph Gonzalez, Michael Jordan, and Ion Stoica.
\newblock Cilantro: {Performance-Aware} resource allocation for general
  objectives via online feedback.
\newblock In {\em 17th USENIX Symposium on Operating Systems Design and
  Implementation (OSDI 23)}, pages 623--643, Boston, MA, July 2023. USENIX
  Association.

\bibitem{veritas}
Chandan Bothra, Jianfei Gao, Sanjay Rao, and Bruno Ribeiro.
\newblock Veritas: Answering causal queries from video streaming traces.
\newblock In {\em Proceedings of the ACM SIGCOMM 2023 Conference}, ACM SIGCOMM
  '23, pages 738--753, New York, NY, USA, 2023. Association for Computing
  Machinery.

\bibitem{charnes-cooper-1962}
Abraham Charnes and William~W. Cooper.
\newblock Programming with linear fractional functionals.
\newblock {\em Naval Research Logistics Quarterly}, 9(3--4):181--186, 1962.

\bibitem{diffprov}
Ang Chen, Yang Wu, Andreas Haeberlen, Wenchao Zhou, and Boon~Thau Loo.
\newblock The good, the bad, and the differences: {B}etter network diagnostics
  with differential provenance.
\newblock In {\em 2016 ACM SIGCOMM Conference (SIGCOMM '16)}, August 2016.

\bibitem{chvatal1979greedy}
V.~Chvatal.
\newblock A greedy heuristic for the set-covering problem.
\newblock {\em Math. Oper. Res.}, 4(3):233–235, August 1979.
\newblock \url{https://doi.org/10.1287/moor.4.3.233}.

\bibitem{k8s_configure_probes}
{Configuring Kubernetes}.
\newblock {Liveness, Readiness and Startup Probes}.
\newblock
  \url{https://kubernetes.io/docs/tasks/configure-pod-container/configure-liveness-readiness-startup-probes/}.
\newblock Accessed: 2025-07-08.

\bibitem{datadog-containers-serverless-2025}
{Datadog}.
\newblock State of containers and serverless.
\newblock \url{https://www.datadoghq.com/state-of-containers-and-serverless/},
  2026.
\newblock Accessed: 2026-09-03.

\bibitem{cloudlab}
Dmitry Duplyakin, Robert Ricci, Aleksander Maricq, Gary Wong, Jonathon Duerig,
  Eric Eide, Leigh Stoller, Mike Hibler, David Johnson, Kirk Webb, Aditya
  Akella, Kuangching Wang, Glenn Ricart, Larry Landweber, Chip Elliott, Michael
  Zink, Emmanuel Cecchet, Snigdhaswin Kar, and Prabodh Mishra.
\newblock The design and operation of {CloudLab}.
\newblock In {\em Proceedings of the {USENIX} Annual Technical Conference
  (ATC)}, pages 1--14, July 2019.

\bibitem{sage}
Yu~Gan, Mingyu Liang, Sundar Dev, David Lo, and Christina Delimitrou.
\newblock Sage: Practical and scalable ml-driven performance debugging in
  microservices.
\newblock In {\em Proceedings of the 26th ACM International Conference on
  Architectural Support for Programming Languages and Operating Systems},
  ASPLOS '21, pages 135--151, New York, NY, USA, 2021. Association for
  Computing Machinery.

\bibitem{deathstarbench}
Yu~Gan, Yanqi Zhang, Dailun Cheng, Ankitha Shetty, Priyal Rathi, Nayan Katarki,
  Ariana Bruno, Justin Hu, Brian Ritchken, Brendon Jackson, Kelvin Hu, Meghna
  Pancholi, Yuan He, Brett Clancy, Chris Colen, Fukang Wen, Catherine Leung,
  Siyuan Wang, Leon Zaruvinsky, Mateo Espinosa, Rick Lin, Zhongling Liu, Jake
  Padilla, and Christina Delimitrou.
\newblock An open-source benchmark suite for microservices and their
  hardware-software implications for cloud \& edge systems.
\newblock In {\em Proceedings of the Twenty-Fourth International Conference on
  Architectural Support for Programming Languages and Operating Systems},
  ASPLOS '19, page 3–18, New York, NY, USA, 2019. Association for Computing
  Machinery.

\bibitem{virelay}
Saksham Goel, Benjamin Mikek, Jehad Aly, Venkat Arun, Ahmed Saeed, and Aditya
  Akella.
\newblock Quantitative verification of scheduling heuristics.
\newblock arXiv, 2023.

\bibitem{gcp-autoscaler}
{Google Cloud}.
\newblock Autoscaling groups of instances.
\newblock \url{https://cloud.google.com/compute/docs/autoscaler}, 2024.
\newblock Accessed: 2025-06-20.

\bibitem{gke-metric-autoscaling}
{Google Cloud}.
\newblock About autoscaling workloads based on metrics.
\newblock
  \url{https://docs.cloud.google.com/kubernetes-engine/docs/concepts/custom-and-external-metrics},
  2026.
\newblock Google Kubernetes Engine documentation, last updated 2026-03-26,
  accessed 2026-03-30.

\bibitem{online-boutique-github}
GoogleCloudPlatform.
\newblock Online {Boutique}: A cloud-native microservices demo application.
\newblock \url{https://github.com/GoogleCloudPlatform/microservices-demo},
  2020.

\bibitem{harvestcontainers}
Adam Hall, Anirudh Sarma, Esha Choukse, Umakishore Ramachandran, and Sameh
  Elnikety.
\newblock Harvesting spare {CPU} resources in container systems.
\newblock In {\em 23rd USENIX Symposium on Networked Systems Design and
  Implementation (NSDI 26)}, pages 2323--2337, Renton, WA, May 2026. USENIX
  Association.

\bibitem{murphy}
Vipul Harsh, Wenxuan Zhou, Sachin Ashok, Radhika~Niranjan Mysore, Brighten
  Godfrey, and Sujata Banerjee.
\newblock Murphy: Performance diagnosis of distributed cloud applications.
\newblock In {\em Proceedings of the ACM SIGCOMM 2023 Conference}, pages
  438--451, 2023.

\bibitem{perf-contract-nf}
Rishabh Iyer, Luis Pedrosa, Arseniy Zaostrovnykh, Solal Pirelli, Katerina
  Argyraki, and George Candea.
\newblock Performance contracts for software network functions.
\newblock In {\em 16th USENIX Symposium on Networked Systems Design and
  Implementation (NSDI 19)}, pages 517--530, 2019.

\bibitem{webperf}
Yurong Jiang, Lenin~Ravindranath Sivalingam, Suman Nath, and Ramesh Govindan.
\newblock Webperf: Evaluating "what-if" scenarios for cloud-hosted web
  applications.
\newblock Technical Report MSR-TR-2016-16, June 2016.

\bibitem{joblib}
{Joblib Development Team}.
\newblock Joblib: running python functions as pipeline jobs, 2020.
\newblock \url{https://joblib.readthedocs.io/}.

\bibitem{google-sre}
Chris Jones, John Wilkes, Niall Murphy, and Cody Smith.
\newblock Service level objectives.
\newblock In Betsy Beyer, Chris Jones, Jennifer Petoff, and Niall~Richard
  Murphy, editors, {\em Site Reliability Engineering: How Google Runs
  Production Systems}, chapter~4, pages 37--48. O'Reilly Media, 2016.

\bibitem{xplain}
Pantea Karimi, Solal Pirelli, Siva Kesava~Reddy Kakarla, Ryan Beckett, Santiago
  Segarra, Beibin Li, Pooria Namyar, and Behnaz Arzani.
\newblock Towards safer heuristics with xplain.
\newblock In {\em Proceedings of the 23rd ACM Workshop on Hot Topics in
  Networks}, pages 68--76, 2024.

\bibitem{worst-case-llm-heuristics}
Pantea Karimi, Dany Rouhana, Pooria Namyar, Siva Kesava~Reddy Kakarla, Venkat
  Arun, and Behnaz Arzani.
\newblock Robust heuristic algorithm design with llms.
\newblock {\em arXiv preprint arXiv:2510.08755}, 2025.

\bibitem{karp1972reducibility}
Richard~M. Karp.
\newblock Reducibility among combinatorial problems.
\newblock In Raymond~E. Miller and James~W. Thatcher, editors, {\em Complexity
  of Computer Computations}, pages 85--103, New York, 1972. Plenum Press.

\bibitem{knative-kpa}
{Knative Project}.
\newblock Autoscaling concepts.
\newblock
  \url{https://knative.dev/docs/serving/autoscaling/autoscaling-concepts/},
  2024.
\newblock Accessed: 2025-06-20.

\bibitem{k8s-hpa-stabilization-window}
{Kubernetes}.
\newblock Horizontal pod autoscaling.
\newblock
  \url{https://kubernetes.io/docs/concepts/workloads/autoscaling/horizontal-pod-autoscale/}.
\newblock Accessed: 2026-09-17.

\bibitem{k8s-hpa}
{Kubernetes}.
\newblock Horizontal pod autoscaler.
\newblock
  \url{https://kubernetes.io/docs/tasks/run-application/horizontal-pod-autoscale/},
  2024.
\newblock Accessed: 2025-06-20.

\bibitem{kivi}
Bingzhe Liu, Gangmuk Lim, Ryan Beckett, and P.~Brighten Godfrey.
\newblock Kivi: Verification for cluster management.
\newblock In {\em 2024 USENIX Annual Technical Conference (USENIX ATC 24)},
  pages 509--527, Santa Clara, CA, July 2024. USENIX Association.

\bibitem{locust}
{Locust team}.
\newblock {Locust}: A modern load testing framework, ongoing.
\newblock \url{https://locust.io/}.

\bibitem{alibaba-2021}
Shutian Luo, Huanle Xu, Chengzhi Lu, Kejiang Ye, Guoyao Xu, Liping Zhang,
  Yu~Ding, Jian He, and Chengzhong Xu.
\newblock Characterizing microservice dependency and performance: Alibaba trace
  analysis.
\newblock In {\em Proceedings of the ACM Symposium on Cloud Computing}, pages
  412--426, 2021.

\bibitem{erms}
Shutian Luo, Huanle Xu, Kejiang Ye, Guoyao Xu, Liping Zhang, Jian He, Guodong
  Yang, and Chengzhong Xu.
\newblock Erms: Efficient resource management for shared microservices with sla
  guarantees.
\newblock In {\em Proceedings of the 28th ACM International Conference on
  Architectural Support for Programming Languages and Operating Systems, Volume
  1}, ASPLOS 2023, page 62–77, New York, NY, USA, 2022. Association for
  Computing Machinery.

\bibitem{metaopt}
Pooria Namyar, Behnaz Arzani, Ryan Beckett, Santiago Segarra, Himanshu Raj,
  Umesh Krishnaswamy, Ramesh Govindan, and Srikanth Kandula.
\newblock Finding adversarial inputs for heuristics using multi-level
  optimization.
\newblock In {\em 21st USENIX Symposium on Networked Systems Design and
  Implementation (NSDI 24)}, pages 927--949, 2024.

\bibitem{oracle-jvm-tuning}
{Oracle}.
\newblock {JVM} tuning.
\newblock
  \url{https://docs.oracle.com/cd/E11035_01/wls100/perform/JVMTuning.html}.
\newblock Oracle WebLogic Server 10.0 Performance and Tuning.

\bibitem{syntra}
Tony Pan, Anup Agarwal, Isil Dillig, and Venkat Arun.
\newblock Syntra: Synthesizing cross-layer controllers for low-latency video
  streaming.
\newblock In {\em USENIX NSDI}, 2026.

\bibitem{graf}
Jinwoo Park, Byungkwon Choi, Chunghan Lee, and Dongsu Han.
\newblock Graf: a graph neural network based proactive resource allocation
  framework for slo-oriented microservices.
\newblock In {\em Proceedings of the 17th International Conference on Emerging
  Networking EXperiments and Technologies}, CoNEXT '21, page 154–167, New
  York, NY, USA, 2021. Association for Computing Machinery.

\bibitem{adversarial-nf}
Luis Pedrosa, Rishabh Iyer, Arseniy Zaostrovnykh, Jonas Fietz, and Katerina
  Argyraki.
\newblock Automated synthesis of adversarial workloads for network functions.
\newblock In {\em Proceedings of the 2018 Conference of the ACM Special
  Interest Group on Data Communication}, pages 372--385, 2018.

\bibitem{thesios}
Phitchaya~Mangpo Phothilimthana, Saurabh Kadekodi, Soroush Ghodrati, Selene
  Moon, and Martin Maas.
\newblock Thesios: Synthesizing accurate counterfactual i/o traces from i/o
  samples.
\newblock In {\em Proceedings of the 29th ACM International Conference on
  Architectural Support for Programming Languages and Operating Systems, Volume
  3}, ASPLOS '24, page 1016–1032, New York, NY, USA, 2024. Association for
  Computing Machinery.

\bibitem{pulp}
{PuLP}.
\newblock Pulp: A linear programming toolkit for python.
\newblock \url{https://coin-or.github.io/pulp/}.
\newblock Accessed: 2025-07-08.

\bibitem{firm-osdi}
Haoran Qiu, Subho~S. Banerjee, Saurabh Jha, Zbigniew~T. Kalbarczyk, and
  Ravishankar~K. Iyer.
\newblock {FIRM}: An intelligent fine-grained resource management framework for
  {SLO-Oriented} microservices.
\newblock In {\em 14th USENIX Symposium on Operating Systems Design and
  Implementation (OSDI 20)}, pages 805--825. USENIX Association, November 2020.

\bibitem{autopilot}
Krzysztof Rzadca, Pawel Findeisen, Jacek Swiderski, Przemyslaw Zych, Przemyslaw
  Broniek, Jarek Kusmierek, Pawel Nowak, Beata Strack, Piotr Witusowski, Steven
  Hand, and John Wilkes.
\newblock Autopilot: workload autoscaling at google.
\newblock In {\em Proceedings of the Fifteenth European Conference on Computer
  Systems}, EuroSys '20, New York, NY, USA, 2020. Association for Computing
  Machinery.

\bibitem{erlang}
Vighnesh Sachidananda and Anirudh Sivaraman.
\newblock Erlang: Application-aware autoscaling for cloud microservices.
\newblock In {\em Proceedings of the Nineteenth European Conference on Computer
  Systems}, EuroSys '24, page 888–923, New York, NY, USA, 2024. Association
  for Computing Machinery.

\bibitem{percs}
Divyanshu Saxena, Gaurav Vipat, Jiaxin Lin, Jingbo Wang, Isil Dillig, Sanjay
  Shakkottai, and Aditya Akella.
\newblock Towards performance robustness for microservices.
\newblock In {\em 23rd USENIX Symposium on Networked Systems Design and
  Implementation (NSDI 26)}, pages 93--112, Renton, WA, May 2026. USENIX
  Association.

\bibitem{jaeger_uber_2017}
Yuri Shkuro.
\newblock Evolving distributed tracing at uber engineering.
\newblock \url{https://www.uber.com/blog/distributed-tracing/}, 2017.
\newblock Accessed: 2026-04-01.

\bibitem{oppertune}
Gagan Somashekar, Karan Tandon, Anush Kini, Chieh-Chun Chang, Petr Husak,
  Ranjita Bhagwan, Mayukh Das, Anshul Gandhi, and Nagarajan Natarajan.
\newblock {OPPerTune}: {Post-Deployment} configuration tuning of services made
  easy.
\newblock In {\em 21st USENIX Symposium on Networked Systems Design and
  Implementation (NSDI 24)}, pages 1101--1120, Santa Clara, CA, April 2024.
  USENIX Association.

\bibitem{anvil}
Xudong Sun, Wenjie Ma, Jiawei~Tyler Gu, Zicheng Ma, Tej Chajed, Jon Howell,
  Andrea Lattuada, Oded Padon, Lalith Suresh, Adriana Szekeres, and Tianyin Xu.
\newblock Anvil: Verifying liveness of cluster management controllers.
\newblock In {\em 18th USENIX Symposium on Operating Systems Design and
  Implementation (OSDI 24)}, pages 649--666, Santa Clara, CA, July 2024. USENIX
  Association.

\bibitem{google-sre-workbook-impl-slos}
Steven Thurgood, David Ferguson, Alex Hidalgo, and Betsy Beyer.
\newblock Implementing {SLOs}, 2018.
\newblock \url{https://sre.google/workbook/implementing-slos/}.

\bibitem{zipkin_twitter_2012}
{Twitter, Inc.}
\newblock Zipkin: A distributed tracing system.
\newblock \url{https://zipkin.io/}, 2012.
\newblock Accessed: 2026-04-01.

\bibitem{teastore}
J{\'o}akim von Kistowski, Simon Eismann, Norbert Schmitt, Andr{\'e} Bauer,
  Johannes Grohmann, and Samuel Kounev.
\newblock {TeaStore: A Micro-Service Reference Application for Benchmarking,
  Modeling and Resource Management Research}.
\newblock In {\em Proceedings of the 26th IEEE International Symposium on the
  Modelling, Analysis, and Simulation of Computer and Telecommunication
  Systems}, MASCOTS '18, September 2018.

\bibitem{autothrottle}
Zibo Wang, Pinghe Li, Chieh-Jan~Mike Liang, Feng Wu, and Francis~Y. Yan.
\newblock Autothrottle: A practical {Bi-Level} approach to resource management
  for {SLO-Targeted} microservices.
\newblock In {\em 21st USENIX Symposium on Networked Systems Design and
  Implementation (NSDI 24)}, pages 149--165, Santa Clara, CA, April 2024.
  USENIX Association.

\bibitem{zeno}
Yang Wu, Ang Chen, and Linh Thi~Xuan Phan.
\newblock Zeno: Diagnosing performance problems with temporal provenance.
\newblock In {\em 16th USENIX Symposium on Networked Systems Design and
  Implementation (NSDI 19)}, pages 395--420, Boston, MA, February 2019. USENIX
  Association.

\bibitem{negative-provenance}
Yang Wu, Mingchen Zhao, Andreas Haeberlen, Wenchao Zhou, and Boon~Thau Loo.
\newblock Diagnosing missing events in distributed systems with negative
  provenance.
\newblock In {\em Proceedings of the 2014 ACM Conference on SIGCOMM}, SIGCOMM
  '14, page 383–394, New York, NY, USA, 2014. Association for Computing
  Machinery.

\bibitem{slowpoke}
Yizheng Xie, Di~Jin, O{\u g}uzhan {\c C}{\"o}lkesen, Vasiliki Kalavri, John
  Liagouris, and Nikos Vasilakis.
\newblock Slowpoke: End-to-end throughput optimization modeling for
  microservice applications.
\newblock In {\em 23rd USENIX Symposium on Networked Systems Design and
  Implementation (NSDI 26)}, pages 631--646, Renton, WA, May 2026. USENIX
  Association.

\bibitem{tiersizer}
Sujay Yadalam, Saarth Deshpande, Divyanshu Saxena, Venkat Arun, Neeraja~J.
  Yadwadkar, and Aditya Akella.
\newblock {TierSizer}: Counterfactual reasoning for {DRAM} sizing in tiered
  memory systems.
\newblock In {\em 4th Workshop on Disruptive Memory Systems ({DIMES})}, 2026.

\bibitem{performal}
Tony~Nuda Zhang, Upamanyu Sharma, and Manos Kapritsos.
\newblock Performal: Formal verification of latency properties for distributed
  systems.
\newblock {\em Proc. ACM Program. Lang.}, 7(PLDI), June 2023.

\bibitem{sinan-delimitrou}
Yanqi Zhang, Weizhe Hua, Zhuangzhuang Zhou, G.~Edward Suh, and Christina
  Delimitrou.
\newblock Sinan: Ml-based and qos-aware resource management for cloud
  microservices.
\newblock In {\em Proceedings of the 26th ACM International Conference on
  Architectural Support for Programming Languages and Operating Systems},
  ASPLOS '21, page 167–181, New York, NY, USA, 2021. Association for
  Computing Machinery.

\bibitem{ursa}
Yanqi Zhang, Zhuangzhuang Zhou, Sameh Elnikety, and Christina Delimitrou.
\newblock { Ursa: Lightweight Resource Management for Cloud-Native
  Microservices }.
\newblock In {\em 2024 IEEE International Symposium on High-Performance
  Computer Architecture (HPCA)}, pages 954--969, Los Alamitos, CA, USA, March
  2024. IEEE Computer Society.

\bibitem{crisp}
Zhizhou Zhang, Murali~Krishna Ramanathan, Prithvi Raj, Abhishek Parwal, Timothy
  Sherwood, and Milind Chabbi.
\newblock {CRISP}: Critical path analysis of {Large-Scale} microservice
  architectures.
\newblock In {\em 2022 USENIX Annual Technical Conference (USENIX ATC 22)},
  pages 655--672, Carlsbad, CA, July 2022. USENIX Association.

\end{thebibliography}

\clearpage

\appendix

\section{Utilization model optimization}\label{appendix:lp-fitting}

Since the objective in Equation~\ref{eqn:num-pods} is not linear in the LP variables $\beta_i$ and $\epsilon_i$, the objective is linear-fractional and can be converted into an LP using a standard Charnes--Cooper transformation~\cite{charnes-cooper-1962}. 

We introduce an auxiliary variable $X_{i}$ with constraints $X_{i} \ge \sum_j \alpha_{i,j}(\rps_j)$, so that the objective becomes $\frac{X_{i}}{T{i} - \beta_{i} - \epsilon_{i}}$. We then define $\lambda_{i} = \frac{1}{T_{i} - \beta_{i} - \epsilon_{i}}$ and scaled variables $\hat{\alpha}_{i,j}(\cdot) = \lambda_{i}\alpha_{i,j}(\cdot)$, $\hat{\beta}_{i} = \lambda_{i}\beta_{i}$, $\hat{\epsilon}_{i} = \lambda_{i}\epsilon_{i}$, and $\hat{X}_{i} = \lambda_{i}X_{i}$. This yields the normalization constraint $T_{i}\lambda_{i} - \hat{\beta}_{i} - \hat{\epsilon}_{i} = 1$ and turns the objective into minimizing $\hat{X}_{i}$. The constraints $X_{i} \ge \sum_j \alpha_{i,j}(\rps_j)$ become $\hat{X}_{i} \ge \sum_j \hat{\alpha}_{i,j}(\rps_j)$, which are linear. The utilization-fit and monotonicity constraints remain linear under the same change of variables. Thus, for each candidate threshold $T_i$, training reduces to solving LPs rather than a nonlinear program. In practice, we solve two LPs in sequence: the first minimizes the transformed cost objective, and the second fixes that optimum while minimizing $\hat{\epsilon}_{i}$ to prefer tighter uncertainty bounds among equally cost-effective solutions.

\section{SLO-safe threshold selection}\label{appendix:slo-algorithm}

We reduce set cover~\cite{karp1972reducibility} to threshold selection. Consider an arbitrary set-cover instance with universe $U=\{u_1,\ldots,u_n\}$ and sets $\mathcal{S}=\{S_1,\ldots,S_d\}$. We construct a threshold-selection instance with one utilization threshold for each set $S_j$ and one SLO-violating observation for each element $u_i$. Thus, each observation has $d$ utilization values, one for each service.

An observation is excluded if, for at least one service, its observed utilization is greater than or equal to that service’s threshold. Choose constants $0\leq a<b<99$. For the observation representing $u_i$, its utilization in dimension $j$ is $b$ if $u_i\in S_j$ and $a$ otherwise. Therefore, lowering threshold $j$ from 99 to $b$ excludes exactly the violations representing the elements in $S_j$. In other words, lowering threshold $j$ has the same effect as selecting set $S_j$.

Define each threshold's additional replica cost as 0 above $b$, 1 at $b$, and $d+1$ below $b$. These costs are non-increasing in the threshold. Lowering all $d$ thresholds to $b$ excludes every observation at an additional cost of $d$, so no optimal solution lowers a threshold below $b$. Any threshold above $b$ can be replaced by 99 without changing its cost or which observations it excludes.

Consequently, lowering $k$ thresholds to $b$ excludes every violation exactly when their corresponding $k$ sets cover $U$, at a cost of $k$ additional replicas. Minimizing this cost therefore finds a minimum set cover. The construction has polynomial size, proving that threshold selection with general monotone costs is NP-hard.

\paragraph{Greedy algorithm.}
Let $R$ be the distinct, componentwise-minimal SLO-violating observations with all utilization values below 99, $T_j$ the finite nonempty set of representative valid thresholds for utilization dimension $j$, and $f_j(t)$ the replica requirement obtained by fitting the utilization model at threshold $t$. For each dimension, thresholds excluding the same observations in $R$ are represented by their largest valid original candidate. Each $f_j$ is non-increasing. For service $s$, let $J_s$ contain its utilization dimensions (CPU and memory). Its replica requirement is
\[
    C_s(\mathbf{v}) = \max_{j \in J_s} f_j(v_j).
\]
We consider service-level replica budgets drawn from
\[
    \mathcal{B}_s =
    \{f_j(t) : j \in J_s,\ t \in T_j\}.
\]
For each budget, we lower that service's thresholds as far as the budget permits. Adapting the greedy set-cover heuristic~\cite{chvatal1979greedy}, we choose the move with the smallest additional replica cost per newly excluded violation, repeating until all violations are excluded (Algorithm~\ref{alg:greedy-threshold-selection}). An observation is excluded when its utilization in any dimension is greater than or equal to that dimension's threshold.

\algrenewcommand\algorithmicrequire{\textbf{Input:}}
\algrenewcommand\algorithmicensure{\textbf{Output:}}

\begin{algorithm}[!t]
\caption{Greedy SLO-safe threshold selection}
\label{alg:greedy-threshold-selection}
\begin{algorithmic}[1]
\Require Violations $R$; threshold sets $T_j$;
replica costs $f_j$; service dimensions $J_s$
\Ensure SLO-safe thresholds $\mathbf{v}$,
or \textbf{infeasible}

\State $v_j \gets \max T_j$ for every dimension $j$
\State $R \gets
    \{\mathbf{x} \in R : \forall j,\ x_j < v_j\}$
\State $\mathcal{B}_s \gets
    \{f_j(t) : j \in J_s,\ t \in T_j\}$
    for every service $s$
\State $C_s(\mathbf{z}) =
    \max_{j \in J_s} f_j(z_j)$
    for every service $s$
    
\While{$R \neq \varnothing$}
    \State $\rho \gets \infty$;
        $\mathit{best} \gets \textbf{none}$

    \ForAll{services $s$}
        \ForAll{$b \in \mathcal{B}_s : b \ge C_s(\mathbf{v})$}
            \State $\mathbf{u} \gets \mathbf{v}$
            \ForAll{$j \in J_s$}
                \State $u_j \gets
                    \min\{t \in T_j :
                    t \le v_j,\ f_j(t) \le b\}$
            \EndFor

            \State $E \gets
                \{\mathbf{x} \in R :
                \exists j \in J_s,\ x_j \ge u_j\}$

            \If{$E \neq \varnothing$}
                \State $\Delta \gets
                    C_s(\mathbf{u}) - C_s(\mathbf{v})$
                \If{$\Delta / |E| < \rho$}
                    \State $\rho \gets \Delta / |E|$
                    \State $\mathit{best} \gets
                        (\mathbf{u}, E)$
                \EndIf
            \EndIf
        \EndFor
    \EndFor

    \If{$\mathit{best} = \textbf{none}$}
        \State \Return \textbf{infeasible}
    \EndIf
    \State $(\mathbf{v}, E) \gets \mathit{best}$
    \State $R \gets R \setminus E$
\EndWhile

\State \Return $\mathbf{v}$
\end{algorithmic}
\end{algorithm}

\section{Evaluation Workloads}\label{appendix:workloads}

\begin{figure}[!htbp]
    \centering
    \includegraphics[width=\columnwidth]{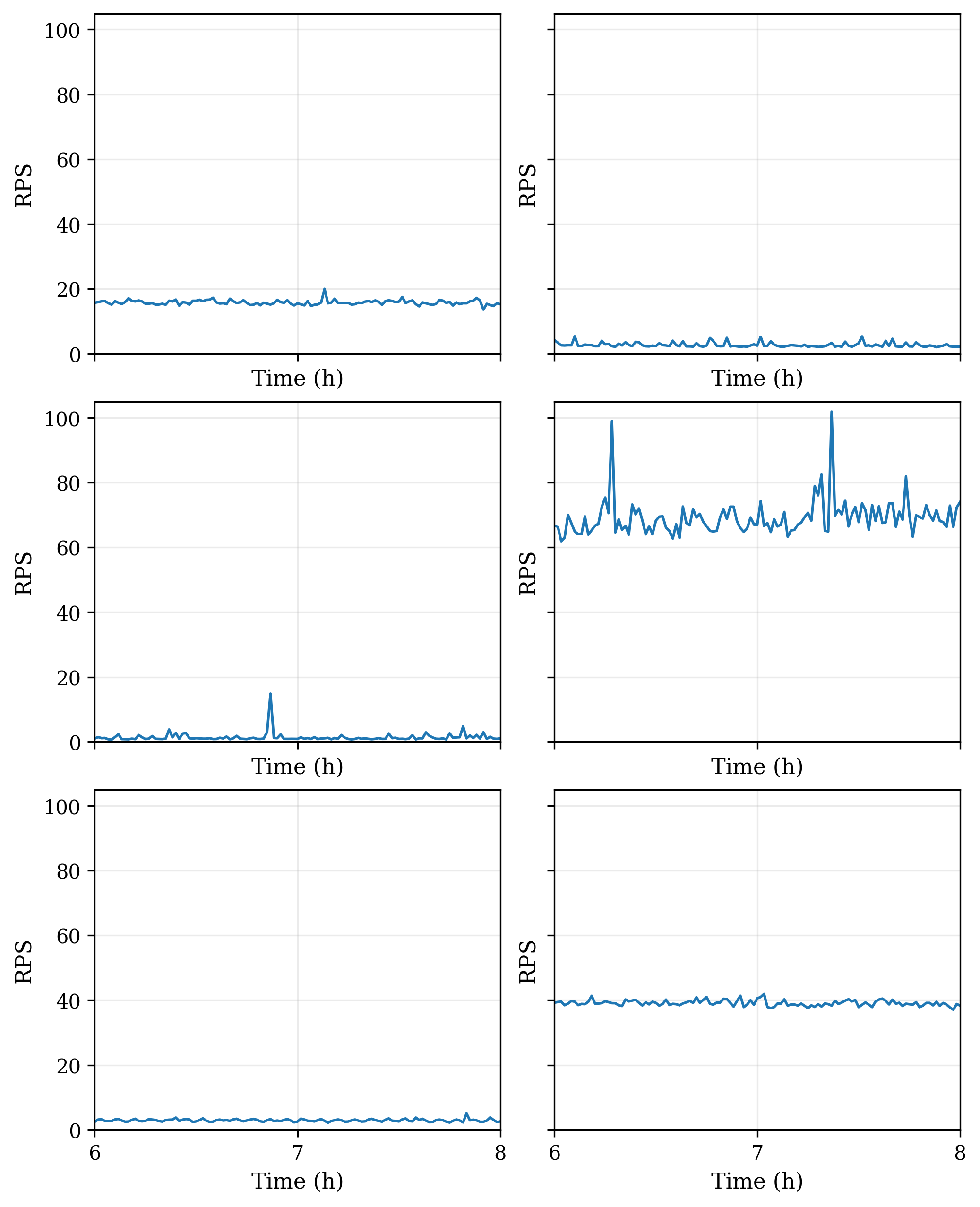}
    \caption{Representative replay workloads used in evaluation. Each panel shows the 6--8\,h evaluation segment of a 12\,h Alibaba-derived request-rate trace. One such trace is assigned to each application endpoint during replay.}
    \label{fig:workloads}
\end{figure}

Figure~\ref{fig:workloads} shows representative 6--8\,h segments from the Alibaba-derived traces used during evaluation. These segments are the intervals on which we evaluate autoscaler behavior after training on the preceding 6 hours. The traces exhibit substantial diversity: some are nearly flat, some show small bursty fluctuations, and others contain sharp spikes or sustained high load. During replay, we assign one such trace to each application endpoint, so different endpoints within the same application may experience qualitatively different demand patterns over the evaluation window.

\FloatBarrier
\FloatBarrier

\section{Evaluating \name on CloudLab}\label{cloudlab-eval}

We run all experiments on a CloudLab~\cite{cloudlab} cluster of r320 machines running Kubernetes with \texttt{kubeadm}. \name's Autoscaler runs on a dedicated node on which Kubernetes pod scheduling is disabled, so the autoscaler itself does not contend for resources the application needs.

\definecolor{hpaColor}{HTML}{3B82F6}
\definecolor{erlangColor}{HTML}{F97316}
\definecolor{whatifColor}{HTML}{16A34A}

\pgfplotstableread[col sep=comma]
    {data/autoscaler-numbers.csv}\latencydata

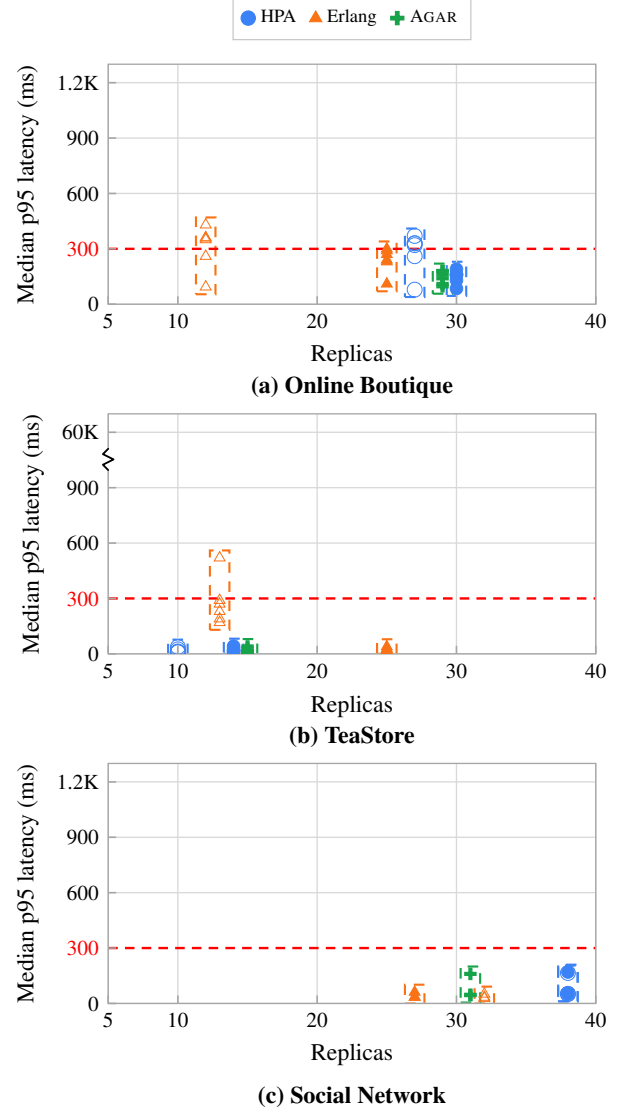
\begin{figure}[!t]
\centering
\begin{tikzpicture}
\begin{groupplot}[
    group style={group size=1 by 3,vertical sep=1.45cm},
    scale only axis,
    width=6.45cm,
    height=3.175cm,
    xmin=5, xmax=40,
    ymin=0, ymax=1300,
    xtick={5,10,20,30,40,50},
    xticklabels={5,10,20,30,40,50},
    ytick={0,600,900,1200},
    yticklabels={0,600,900,1.2K},
    xlabel={Replicas},
    ylabel={Median p95 latency (ms)},
    label style={font=\small},
    tick label style={font=\footnotesize},
    tick align=inside,
    xtick pos=bottom,
    ytick pos=left,
    major tick length=3pt,
    tick style={draw=black!35,line width=0.6pt},
    axis line style={draw=black!35,line width=0.5pt},
    grid=major,
    grid style={draw=black!15,line width=0.5pt},
    extra y ticks={300},
    extra y tick labels={300},
    extra y tick style={
        grid=major,
        grid style={
            draw=red, line width=0.9pt,
            dash pattern=on 3.6pt off 2.7pt
        },
        tick label style={text=red}
    },
    set layers=standard,
    clip mode=individual,
    clip marker paths=true
]

\nextgroupplot[
    xlabel style={
        at={(axis description cs:0.5,0)},
        anchor=north,
        yshift=-15pt,
        inner sep=0pt
    }
]

\PlotSeriesWithBox{cloudlab}{OB}{HPA-Unfixed}
    {hpaColor}{o}%
\PlotSeriesWithBox{cloudlab}{OB}{HPA-Fixed}
    {hpaColor}{*}%

\PlotSeriesWithBox{cloudlab}{OB}{Erlang-Unfixed}
    {erlangColor}{triangle}%
\PlotSeriesWithBox{cloudlab}{OB}{Erlang-Fixed}
    {erlangColor}{triangle*}%

\PlotSeriesWithBox{cloudlab}{OB}{WhatIf}
    {whatifColor}{filled plus}%

\pgfplotsset{
    legend columns=3,
    legend cell align=left,
    legend style={
        at={(0.5,1.10)},
        anchor=south,
        font=\scriptsize,
        draw=black!25,
        fill=white,
        inner xsep=3pt,
        inner ysep=3pt,
        /tikz/every even column/.append style={
            column sep=3pt
        }
    }
}

\addlegendimage{
    only marks,
    mark=*,
    mark size=2.7pt,
    mark options={solid,draw=hpaColor,fill=hpaColor}
}
\addlegendentry{HPA}

\addlegendimage{
    only marks,
    mark=triangle*,
    mark size=2.7pt,
    mark options={solid,draw=erlangColor,fill=erlangColor}
}
\addlegendentry{Erlang}

\addlegendimage{
    only marks,
    mark=filled plus,
    mark size=2.7pt,
    mark options={solid,draw=whatifColor,fill=whatifColor}
}
\addlegendentry{\name}

\nextgroupplot[
    yticklabels={0,600,900,{60K}}
]
\PlotSeriesWithBox{cloudlab}{TS}{HPA-Unfixed}
    {hpaColor}{o}%
\PlotSeriesWithBox{cloudlab}{TS}{HPA-Fixed}
    {hpaColor}{*}%

\PlotSeriesWithBox{cloudlab}{TS}{Erlang-Unfixed}
    {erlangColor}{triangle}%
\PlotSeriesWithBox{cloudlab}{TS}{Erlang-Fixed}
    {erlangColor}{triangle*}%

\PlotSeriesWithBox{cloudlab}{TS}{WhatIf}
    {whatifColor}{filled plus}%

\nextgroupplot[
    xlabel style={
        at={(axis description cs:0.5,0)},
        anchor=north,
        yshift=-15pt,
        inner sep=0pt
    }
]

\PlotSeriesWithBox{cloudlab}{SN}{HPA-Unfixed}
    {hpaColor}{o}%
\PlotSeriesWithBox{cloudlab}{SN}{HPA-Fixed}
    {hpaColor}{*}%

\PlotSeriesWithBox{cloudlab}{SN}{Erlang-Unfixed}
    {erlangColor}{triangle}%
\PlotSeriesWithBox{cloudlab}{SN}{Erlang-Fixed}
    {erlangColor}{triangle*}%

\PlotSeriesWithBox{cloudlab}{SN}{WhatIf}
    {whatifColor}{filled plus}%

\end{groupplot}

\coordinate (tsbreak) at
    ($(group c1r2.south west)!0.81!
      (group c1r2.north west)$);

\draw[
    overlay,
    white,
    line width=3pt
]
    ($(tsbreak)+(0,-4pt)$) --
    ($(tsbreak)+(0,4pt)$);

\draw[
    overlay,
    black,
    line width=0.7pt,
    line cap=round,
    line join=round
]
    ($(tsbreak)+(0,-4pt)$) --
    ($(tsbreak)+(-2pt,-2pt)$) --
    ($(tsbreak)+(2pt,0pt)$) --
    ($(tsbreak)+(-2pt,2pt)$) --
    ($(tsbreak)+(0,4pt)$);

\node[
    font=\small\bfseries,
    anchor=south,
    inner sep=0pt,
    outer sep=0pt
]
    at ([yshift=6pt]group c1r2.north)
    {(a) Online Boutique};

\node[
    font=\small\bfseries,
    anchor=south,
    inner sep=0pt,
    outer sep=0pt
]
    at ([yshift=6pt]group c1r3.north)
    {(b) TeaStore};

\node[font=\small\bfseries]
    at ([yshift=-35pt]group c1r3.south)
    {(c) Social Network};

\end{tikzpicture}

\caption{Per-endpoint median p95 latency when using HPA, Erlang, and \name on CloudLab. The horizontal dashed line marks the 300\,ms \textcolor{red!70!black}{SLO}. Hollow markers indicate unfixed configurations (described in \S\ref{subsec:debugger}). For each application, the boundary boxes mark the average replica count for each autoscaler configuration.}
\label{fig:autoscaler-latencies-cloudlab}

\end{figure}

\section{Effect of Host Interference on \name's Models}\label{model-param-interference-changes}

\pgfplotstableread[col sep=comma]{data/interference-model-params.csv}\interferenceparams

\newcommand{\InterferenceTable}[3]{%
\begin{table*}[!t]
\centering
\small
\renewcommand{\arraystretch}{0.88}
\setlength{\aboverulesep}{0.25ex}
\setlength{\belowrulesep}{0.25ex}
\pgfplotstabletypeset[
    empty cells with={--},
    columns={
        service,
        metric,
        beta_dedicated,
        beta_shared,
        epsilon_dedicated,
        epsilon_shared,
        threshold_dedicated,
        threshold_shared
    },
    columns/service/.style={
    string type,
    column name={\textbf{Service}},
        postproc cell content/.append code={
            \ifodd\pgfplotstablerow
                \pgfkeyssetvalue{/pgfplots/table/@cell content}{}
            \else
                \pgfkeysgetvalue{/pgfplots/table/@cell content}\servicecell
                \edef\temp{
                    \noexpand\pgfkeyssetvalue
                    {/pgfplots/table/@cell content}
                    {\noexpand\multirow{2}{*}{\unexpanded\expandafter{\servicecell}}}
                }
                \temp
            \fi
        }
    },
    columns/metric/.style={
        string type,
        column name={\textbf{Metric}}
    },
    columns/beta_dedicated/.style={
        column name={\textbf{$\beta$ (Dedicated)}},
        fixed,
        fixed zerofill,
        precision=1
    },
    columns/beta_shared/.style={
        column name={\textbf{$\beta$ (Shared)}},
        fixed,
        fixed zerofill,
        precision=1
    },
    columns/epsilon_dedicated/.style={
        column name={\textbf{$\epsilon$ (Dedicated)}},
        fixed,
        fixed zerofill,
        precision=1
    },
    columns/epsilon_shared/.style={
        column name={\textbf{$\epsilon$ (Shared)}},
        fixed,
        fixed zerofill,
        precision=1
    },
    columns/threshold_dedicated/.style={
        column name={\textbf{$T$ (Dedicated)}},
        fixed,
        fixed zerofill,
        precision=1
    },
    columns/threshold_shared/.style={
        column name={\textbf{$T$ (Shared)}},
        fixed,
        fixed zerofill,
        precision=1
    },
    row predicate/.code={
        \pgfplotstablegetelem{##1}{app}\of{\interferenceparams}
        \edef\rowapp{\pgfplotsretval}
        \ifdefstring{\rowapp}{#1}{}{
            \pgfplotstableuserowfalse
        }
    },
    every head row/.style={
        before row=\toprule,
        after row=\midrule
    },
    every odd row/.style={
        after row=\midrule
    },
    every even row/.style={
        after row=\cmidrule(lr){2-8}
    },
    every last row/.style={
        after row=\bottomrule
    },
    col sep=comma
]{\interferenceparams}
\caption{#2}
\label{#3}
\end{table*}
}

\InterferenceTable
    {OB}
    {Learned utilization-model parameters for Online Boutique on EC2 \texttt{t3.2xlarge} instances with dedicated and shared tenancy.}
    {tab:interference-params-ob}

\vspace{50pt}

\InterferenceTable
    {TS}
    {Learned utilization-model parameters for TeaStore on EC2 \texttt{t3.2xlarge} instances with dedicated and shared tenancy.}
    {tab:interference-params-ts}

\InterferenceTable
    {SN}
    {Learned utilization-model parameters for Social Network on EC2 \texttt{t3.2xlarge} instances with dedicated and shared tenancy.}
    {tab:interference-params-sn}

To examine how host-level interference is reflected in \name's learned models, Tables~\ref{tab:interference-params-ob}–\ref{tab:interference-params-sn} compare the fitted utilization-model parameters on dedicated and shared hosts for a 300\,ms p95 latency SLO. Differences in $\beta$ capture shifts in baseline utilization, while differences in $\epsilon$ capture changes in unexplained variability; threshold changes reflect corresponding changes in the utilization regions deemed SLO-safe.

\end{document}